\documentclass[floatfix,twocolumn,showpacs,preprintnumbers,amsmath,amssymb,prl,superscriptaddress,longbibliography]{revtex4-1}
\usepackage{color}
\usepackage[usenames,dvipsnames,svgnames,table]{xcolor}
\usepackage[colorlinks=true,linkcolor=blue,urlcolor=blue,citecolor=blue]{hyperref}

\usepackage{mathtools}
\usepackage{graphicx}
\usepackage{dcolumn}
\usepackage{array}
\usepackage{lipsum}
\usepackage{bm}
\usepackage{subfigure}
\usepackage{amssymb}
\usepackage{multirow}
\usepackage{tabularx}
\usepackage{amsmath}
\usepackage{braket}
\graphicspath{{plots/}}

\begin{document}
\title{
Nonlinear Electronic Density Response in Warm Dense Matter
}

\author{Tobias Dornheim}
\email{t.dornheim@hzdr.de}

\affiliation{Center for Advanced Systems Understanding (CASUS), G\"orlitz, Germany}

\author{Jan Vorberger}

\affiliation{Helmholtz-Zentrum Dresden-Rossendorf, Bautzner Landstra{\ss}e 400, D-01328 Dresden, Germany}

\author{Michael Bonitz}

\affiliation{Institut f\"ur Theoretische Physik und Astrophysik, Christian-Albrechts-Universit\"at zu Kiel, Leibnizstra{\ss}e 15, D-24098 Kiel, Germany}

\begin{abstract}
Warm dense matter (WDM)---an extreme state with high temperatures and densities that occurs e.g. in astrophysical objects---constitutes one of the most active fields in plasma physics and materials science. These conditions can be realized in the lab by shock compression or laser excitation, and the most accurate experimental diagnostics is achieved with lasers and free electron lasers which is theoretically modeled using linear response theory. Here, we present first \textit{ab initio} path integral Monte Carlo results for the nonlinear density response of correlated electrons in WDM and show that for many situations of experimental relevance nonlinear effects cannot be neglected.
%
\end{abstract}

\maketitle

Warm dense matter (WDM) is an exotic state with extreme densities ($r_s=\overline{r}/a_\textnormal{B}\sim1$ with $\overline{r}$ and $a_\textnormal{B}$ being the average interparticle distance and first Bohr radius) and high temperatures ($\theta=k_\textnormal{B}T/E_\textnormal{F}\sim1$ with $T$ and $E_\textnormal{F}$ being the temperature and Fermi energy) that occurs, e.g., in astrophysical objects~\cite{Militzer_2008,Guillot2018, saumon1,becker} and laser-excited solids~\cite{ernstorfer2,ernstorfer},
and on the pathway towards inertial confinement fusion~\cite{hu_ICF}.
Consequently, WDM has emerged as one of the most active frontiers in plasma physics and material science~\cite{fortov_review,siegfried_review,falk_wdm}, and WDM conditions are routinely realized in experiments in large research facilities around the globe (e.g., NIF, SLAC and the European XFEL), see Refs.~\cite{Moses_NIF,LCLS_2016,Tschentscher_2017} for review articles. 

On the other hand, the theoretical description of WDM constitutes a formidable challenge~\cite{new_POP,wdm_book} due to the complicated interplay of 1) Coulomb coupling, 2) thermal excitations, and 3) electronic quantum degeneracy effects. 
Moreover, the bulk of WDM theory assumes a weak response of the electrons to an external perturbation, i.e., they rely on linear response theory (LRT). This assumption enters, for example, in the interpretation of XRTS experiments~\cite{siegfried_review,kraus_xrts}, the characterization of the stopping power in WDM \cite{Cayzac2017}, the construction of effective potentials~\cite{ceperley_potential,zhandos1,zhandos2}, density functional theory (DFT) calculations~\cite{dynamic2,pribram}, and the computation of energy relaxation rates~\cite{transfer1,transfer2,ernstorfer}. 
Consequently, numerous works have been devoted to the description of the density response of electrons both in the ground state~\cite{pines,kugler1,stls_original,vs_original,dynamic_ii,farid,moroni,moroni2,bowen2,cdop} and at finite temperature~\cite{stls,schweng,perrot,stolzmann,stls2,tanaka_hnc,dornheim_pre,groth_jcp,arora}. These efforts have culminated in the recent machine-learning representation~\cite{dornheim_ML}  of the static electronic density response that is based on \textit{ab initio} path integral Monte Carlo (PIMC) simulations~\cite{dornheim_electron_liquid,dornheim_HEDP,review} and covers the entire WDM regime. Moreover, even the dynamic density response can be computed from PIMC simulations~\cite{dornheim_dynamic,dynamic_folgepaper}, and the reported negative dispersion relation of a uniform electron gas (UEG) constitutes an active topic of investigation.

On the other hand, very little is known about the density response of correlated electrons beyond the linear regime. In particular, it is unclear up to which perturbation strength LRT remains accurate. This question becomes increasingly urgent, as free electron lasers become more powerful and peak intensities of up to $I\sim10^{22}$W/cm$^2$ \cite{Fletcher2015} have been reported. 
Furthermore, intense VUV lasers are used to probe WDM~\cite{Zastrau}. 
A particular promising tool are THz lasers~\cite{Ofori_Okai_2018} as they allow for probing the low-frequency end of the density response, short pulse characterization, and streaking \cite{Goulielmakis1267,fruehling_np_09,Kazansky_2019}. Yet, THz field applications might require in many cases a theoretical description beyond LRT, as we indicate below.

In this work, we go beyond linear response theory by carrying out extensive PIMC simulations of a harmonically perturbed electron gas~\cite{dornheim_pre,groth_jcp} (cf.~Eq.~(\ref{eq:hamiltonian}) below) at WDM conditions. This allows us to measure the actual density response of the electrons without any a-priori assumptions (including the fluctuation dissipation theorem) and, thus, to unambiguously characterize the validity range of LRT. In addition, going beyond the linear regime allows us to gauge the systematic errors of LRT as a function of perturbation strength, and to report the first results for the cubic response function $\chi_3(q)$ over the entire relevant wave number range for different densities and temperatures \emph{including all exchange--correlation effects}. 
Therefore, our results provide the basis for a  generalized theory of the electronic density response beyond LRT, extending earlier work for classical plasmas \cite{golden_pra_85,bonitz_prl_12} and moderately coupled quantum plasmas 
\cite{kwong_prl-00,haberland_pre_01}.

Our investigation of the nonlinear density response of electrons in WDM should be relevant for many other fields, and spark similar investigations in other domains as we note that LRT is one of the most successful concepts in physics~\cite{nolting,quantum_theory}. It is of paramount importance in many fields, such as for describing phonons in solid state physics~\cite{baroni1,baroni2}, excitations in systems of ultracold atoms~\cite{ultracold1,ultracold2}, and screening or quasiparticle excitations in plasmas ~\cite{plasma1,plasma2}. Moreover, it has allowed for profound physical insights into, e.g., superfluidity~\cite{cep,ultracold2}, collective excitations~\cite{pines,bonitz_book}, and quantum dynamics~\cite{dynamic1,dynamic2}.

 All PIMC data are available online~\cite{supplement} and can be used to benchmark theoretical models and approximate simulation techniques like DFT. 

\textbf{Results.} We simulate a harmonically perturbed electron gas governed by the Hamiltonian (we assume Hartree atomic units throughout this work)
\begin{eqnarray}\label{eq:hamiltonian}
\hat H = \hat H_\textnormal{UEG} + 2 A \sum_{k=1}^N \textnormal{cos}\left( \hat{\mathbf{r}}_k\cdot\hat{\mathbf{q}} \right)\ ,
\end{eqnarray}
with $\hat H_\textnormal{UEG}$ being the usual (unperturbed) UEG Hamiltonian~\cite{loos,quantum_theory,review}, $A$ being the perturbation amplitude, and the wave vector $\mathbf{q} = 2\pi/L (n_x,n_y,n_z)^T$ (with $n_i\in\mathbb{Z}$, and $L$ being the length of the simulation box). Note that we use a canonical adaption~\cite{mezza} of the worm algorithm by Boninsegni \textit{et al.}~\cite{boninsegni1,boninsegni2} without any assumptions on the nodal structure of the thermal density matrix. Therefore, our simulations are computationally involved due to the fermion sign problem~\cite{troyer,dornheim_sign_problem}, but are exact within the given statistical uncertainty.

To measure the density response, we compute the induced density
\begin{eqnarray}\label{eq:rho}
\rho(\mathbf{q},A) \coloneqq \braket{\hat\rho_\mathbf{q}}_A = \frac{1}{V} \left< \sum_{k=1}^N e^{-i\mathbf{q}\cdot\hat{\mathbf{r}}_k} \right>_A \ , \\ \nonumber 
\end{eqnarray}
where $\braket{\dots}_A$ indicates the expectation value computed from Eq.~(\ref{eq:hamiltonian}).
\begin{figure}\centering
\includegraphics[width=0.415\textwidth]{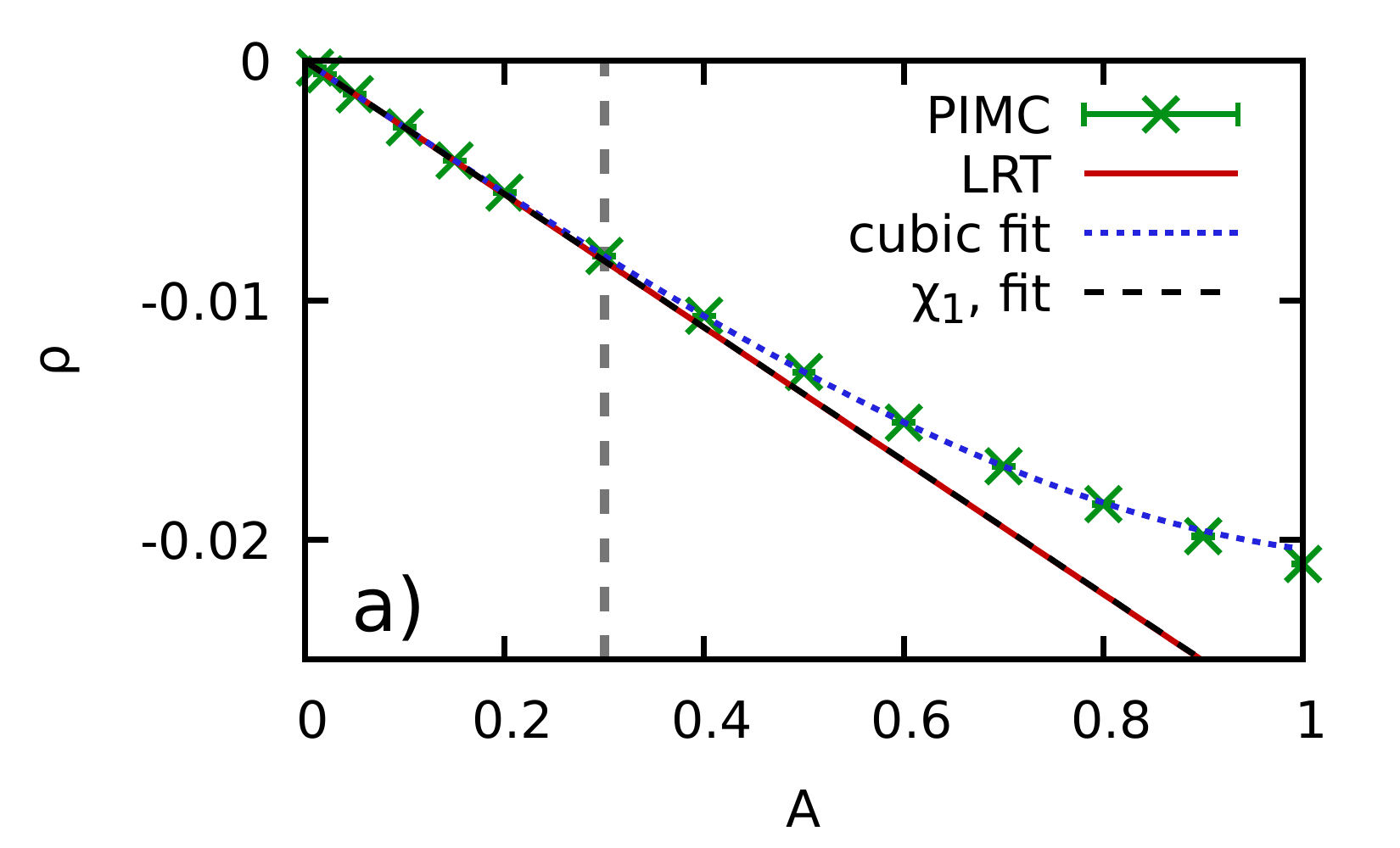}\\ \vspace*{-1.025cm}
\includegraphics[width=0.415\textwidth]{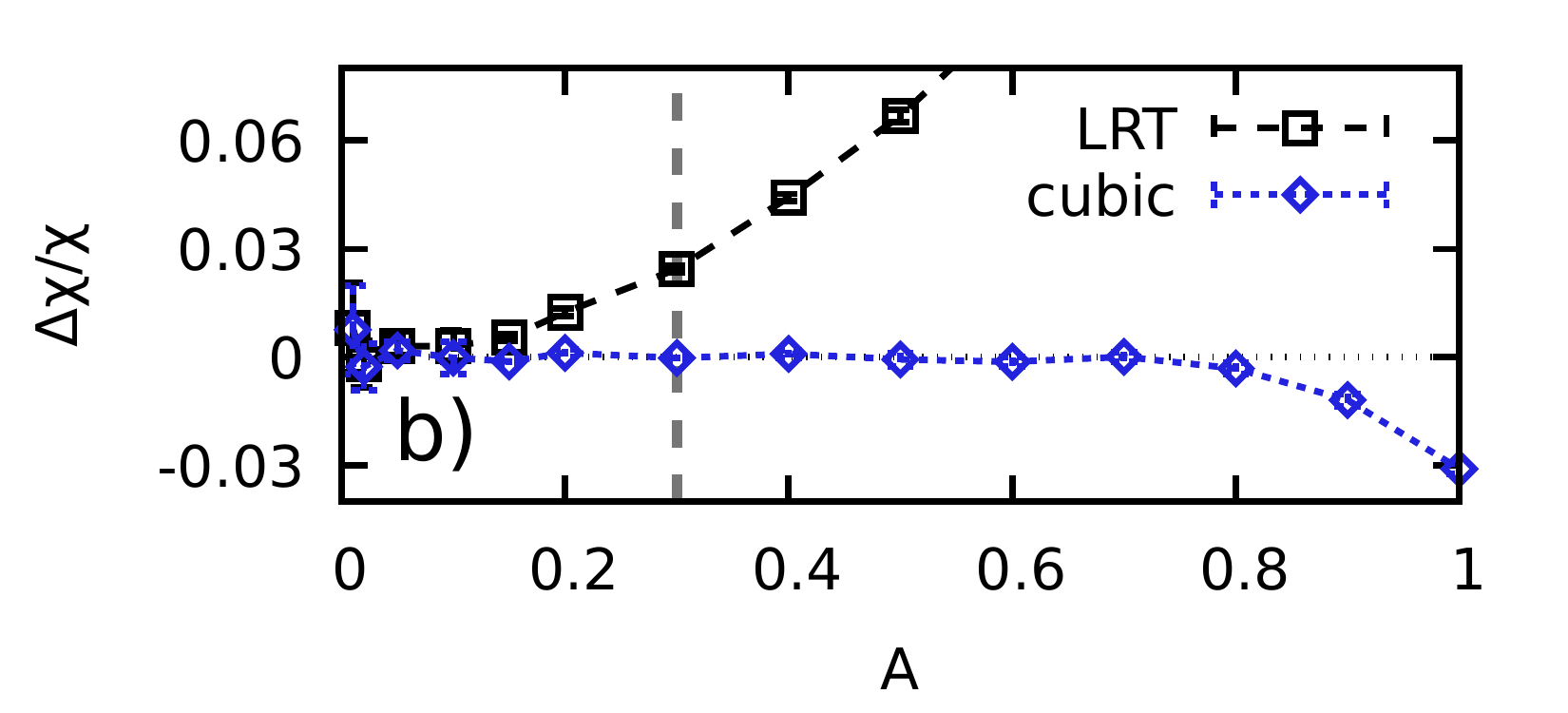}\vspace*{-0.2cm}
\caption{\label{fig:PRL}
Density response of the UEG for $N=14$, $r_s=2$, and $\theta=1$ with $q\approx0.84q_\textnormal{F}$ in dependence of the perturbation amplitude $A$ [cf.~Eq.~(\ref{eq:hamiltonian})]. Panel a) shows the PIMC data for the induced density $\rho$ (green crosses), the prediction from LRT [solid red, cf.~Eq.~(\ref{eq:static_chi})], and a cubic fit [dotted blue, cf.~Eq.~(\ref{eq:fit})] as well as the linear component thereof (dashed black). Panel b) shows the deviation of LRT (black squares) and the cubic fit (blue diamonds) from the PIMC data. The vertical grey dashed line corresponds to the maximum $A$-value that has been included in the fit.
}
\end{figure}  
The PIMC results for Eq.~(\ref{eq:rho}) are shown in Fig.~\ref{fig:PRL}a) as the green crosses for the electron gas with a metallic density ($r_s=2$) at the Fermi temperature, $\theta=1$, for a wave number of $q\approx0.84q_\textnormal{F}$.  
For small $A$, LRT is accurate and it holds
$\rho(\mathbf{q},A) = \chi(\mathbf{q})A$,
and the density response function $\chi(\mathbf{q})$ does not depend on $A$. In this context, we mention that
the linear response function can be computed from a simulation of the unperturbed UEG via the imaginary-time version of the fluctuation--dissipation theorem, which states that
\begin{eqnarray}\label{eq:static_chi}
\chi(\mathbf{q}) = -n\int_0^\beta \textnormal{d}\tau\ F(\mathbf{q},\tau) \quad ,
\end{eqnarray}
with $F(q,\tau)$ being the usual intermediate scattering function~\cite{siegfried_review} evaluated at an imaginary time argument $\tau\in[0,\beta]$, see Ref.~\cite{dornheim_ML} for details. The LRT result for $\rho$ as obtained from Eq.~(\ref{eq:static_chi}) is depicted by the solid red line and is in excellent agreement to the PIMC data for $A\lesssim 0.15$. This can be seen particularly well in Fig.~\ref{fig:PRL}b), where the black squares correspond to the relative deviation between the PIMC data and LRT. We note that LRT systematically overestimates the density response, and the deviation to LRT appears to be parabolic in the depicted $A$-range.

Indeed, it is well known~\cite{moroni,moroni2} that the first term beyond $\chi(q)$ is cubic in $A$ and can be obtained by fitting the PIMC data to
\begin{eqnarray}\label{eq:fit}
\rho(\mathbf{q},A) = \chi_1(q) A + \chi_3(q) A^3 \ ,
\end{eqnarray}
where $\chi_1(q)$ and $\chi_3(q)$ are the free parameters. The results for Eq.~(\ref{eq:fit}) are included in Fig.~\ref{fig:PRL}a) as the dashed blue curve, and exhibit a significantly improved agreement with the PIMC data as compared to LRT. The vertical dashed grey line corresponds to the maximum $A$-value that has been included into the fit, but Eq.~(\ref{eq:fit}) remains accurate for significantly larger perturbation strengths, see also the blue diamonds in panel b). For completeness, we mention that it is, in principle, redundant to obtain $\chi_1(q)$ from the PIMC data, as it is already known from Eq.~(\ref{eq:static_chi}). On the other hand, comparing the two allows to check the consistency of our approach, and the two independent estimations of the LRT function are in perfect agreement with an uncertainty interval of $0.1\%$, see the dashed black line in panel a).

\begin{figure*}\centering
\includegraphics[width=0.353\textwidth]{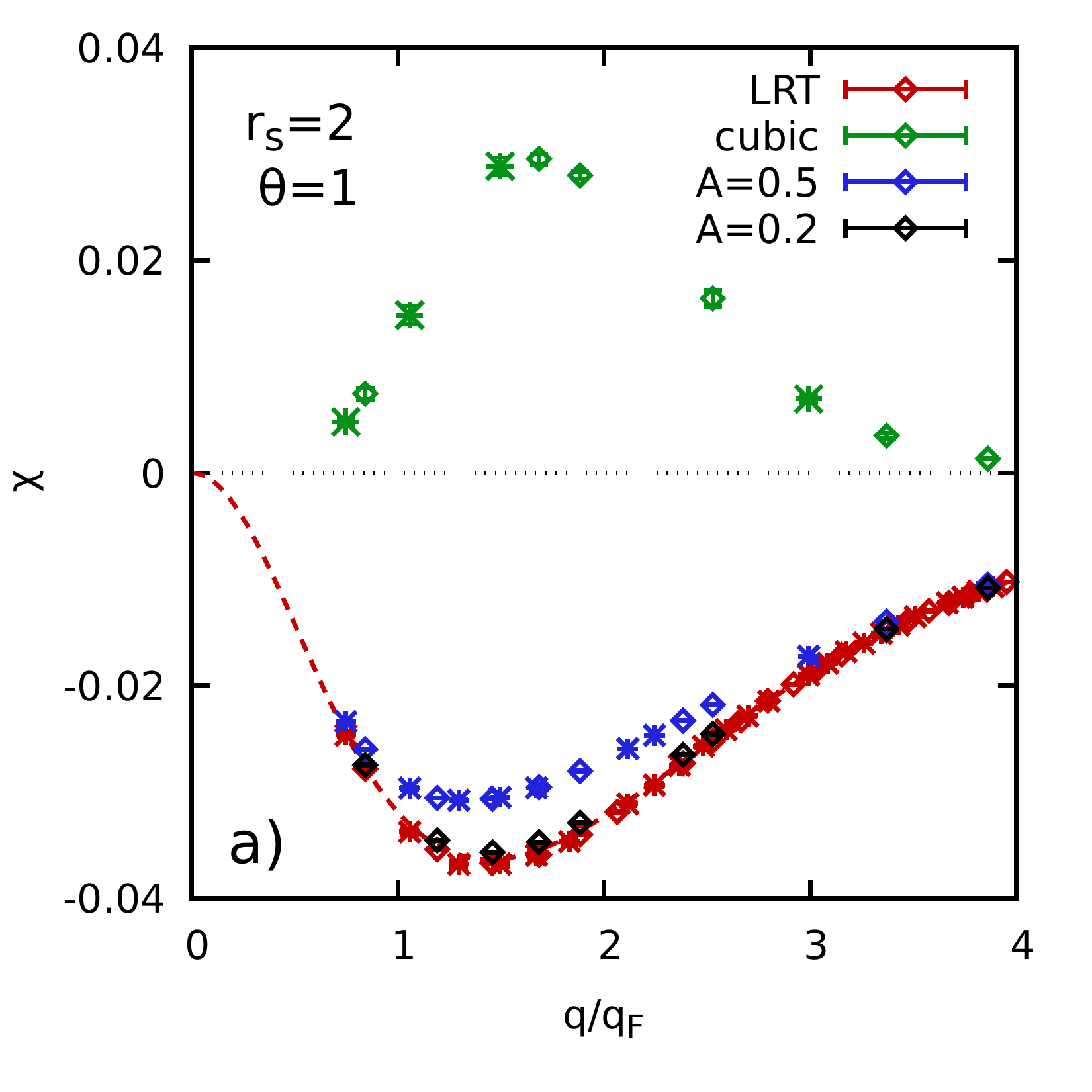}\hspace*{-0.4cm}
\includegraphics[width=0.353\textwidth]{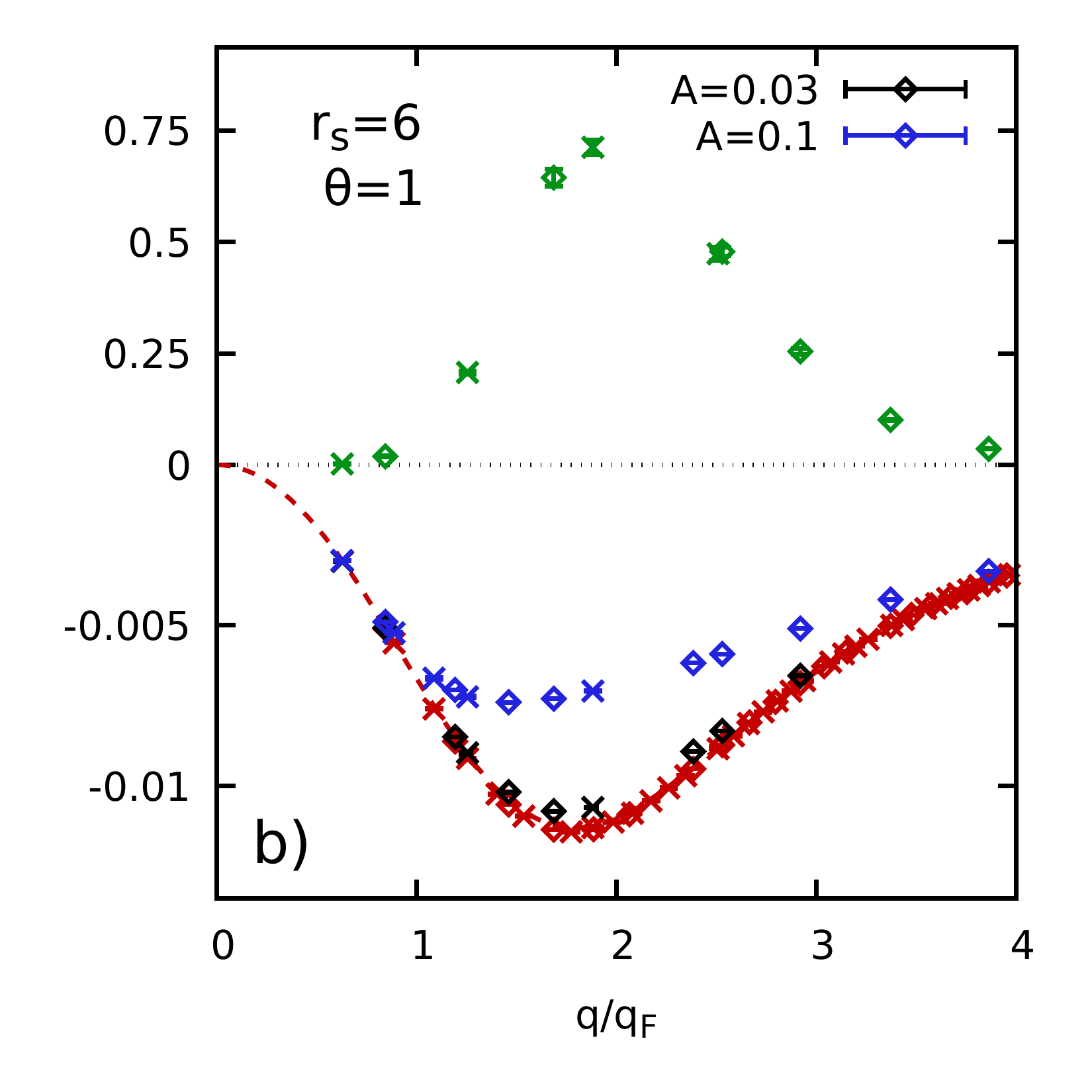}\hspace*{-0.4cm}
\includegraphics[width=0.353\textwidth]{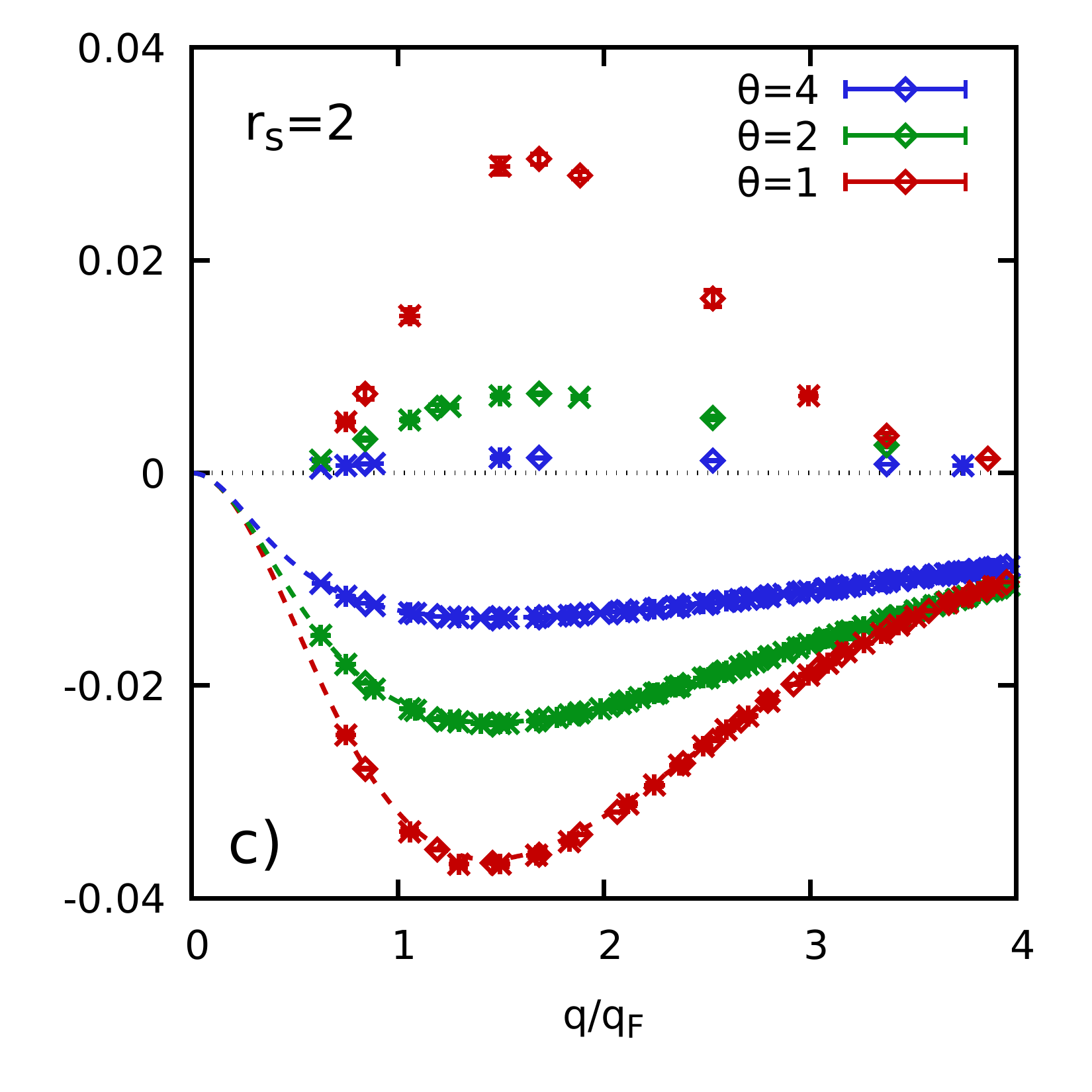}
\caption{\label{fig:overview}
Density response of an electron gas to an external harmonic perturbation at different conditions. Panels a) and b) show results for $\theta=1$ and $r_s=2$ and $r_s=6$, respectively. The top halfs correspond to the cubic response function $\chi_3(q)$ [computed via fits, cf.~Eq.~(\ref{eq:fit})], and the bottom half to the (pseudo-) linear response function $\chi(q)$ from LRT [red, cf.~Eq.~(\ref{eq:static_chi})], and for different perturbation amplitudes [black and blue, cf.~Eq.~(\ref{eq:pseudo})]. Panel c) corresponds to $r_s=2$ for $\theta=4$ (blue), $\theta=2$ (green), and $\theta=1$ (red) and shows $\chi_3$ and $\chi$ (from LRT) in the top and bottom half. The diamonds, stars, and crosses correspond to $N=14, 20,$ and $34$. The dashed curves depict the LRT prediction computed from a recent machine-learning representation~\cite{dornheim_ML} of the static local field correction.
}
\end{figure*}

Let us next investigate the dependence of the response function of warm dense electrons on the wave number $q$. This is shown in Fig.~\ref{fig:overview}a) where the top and bottom half correspond to the cubic and linear response, respectively. The red symbols correspond to the usual LRT function computed from Eq.~(\ref{eq:static_chi}) for $N=14$ (diamonds) and $N=20$ (stars), and the dashed red line to $\chi(q)$ computed in the thermodynamic limit ($N\to\infty$) from the neural-net representation given in Ref.~\cite{dornheim_ML}. We note that they are in good agreement, as finite-size effects are small in this regime~\cite{dornheim_ML}. The black and blue symbols have been obtained from our new PIMC simulations of the perturbed system as
\begin{eqnarray}\label{eq:pseudo}
\chi(\mathbf{q},A) = \frac{\rho(\mathbf{q},A)}{A} \
\end{eqnarray}
such that this \textit{pseudo} response function converges to LRT in the limit of small perturbations,
$\lim_{A\to0} \chi(\mathbf{q},A) = \chi(\mathbf{q})$.
For $A=0.2$ (black symbols), Eq.~(\ref{eq:pseudo}) is in good agreement to the LRT data both for small and large $q$, but systematically deviates around $q\sim q_\textnormal{F}$. For $A=0.5$ (blue symbols), the pseudo response function systematically underestimates the density response over the entire depicted $q$-range, and the discrepancy is again most pronounced for intermediate wave numbers, with a maximum deviation of $\sim20\%$. To more systematically investigate this trend, we have performed extensive $A$-scans such as depicted in Fig.~\ref{fig:PRL} for different $q$-values over the entire relevant wave number range~\cite{supplement}. This has allowed us to obtain the first results for the  cubic response function $\chi_3(q)$, which are shown in the top half of Fig.~\ref{fig:overview}a) as the green data points. As a side note, we mention that a single $\chi_3(q)$ point requires $10-15$ independent PIMC simulations of Eq.~(\ref{eq:hamiltonian}) with different $A$ values for each wave number, which results in a total computation cost of $\mathcal{O}\left(10^7\right)$ CPU hours.

Overall, $\chi_3(q)$ qualitatively somewhat mirrors $\chi(q)$, although with some pronounced differences. First and foremost, we find that no finite-size effects can be resolved within the given error bars, and the results for $N=14$ and $N=20$ exhibit a smooth progression. The main difference is that they are available at different $q$-points, which is a direct consequence of the momentum quantization in the finite simulation cell, see, e.g., Refs.~\cite{dornheim_prl,dornheim_cpp}.
Moreover, $\chi_3(q)$ always has the opposite sign of $\chi(q)$, as the system cannot react arbitrarily strong to the perturbation, and the response eventually saturates. While both the linear and the cubic response function vanish in the large- and small-$q$ limits, this happens significantly sooner for the latter function. Heuristically, this can be understood as follows: for large $q$-values, only single-particle effects contribute to the response, the system as a whole remains hardly affected, and LRT is sufficient; similarly, the response is suppressed by the perfect screening~\cite{kugler_bounds} in the small-$q$ limit.
Lastly, we find that the maximum in $\chi_3(q)$ appears to be slightly shifted to larger $q$-values compared to $\chi(q)$, see also panel b) for the same trend at $r_s=6$. 

In summary, our results predict that nonlinear effects in the electronic density response manifest in an effectively damped response function [cf.~the blue symbols in Fig.~\ref{fig:overview}a)], with a maximum that is shifted to smaller wave numbers.

Let us next investigate the dependence of the cubic response on the density parameter $r_s$. To this end, we repeat our previous study for $r_s=6$, and the results are shown in Fig.~\ref{fig:overview}b) for $\theta=1$. While such low densities are not typical for WDM applications, they can be realized experimentally in hydrogen jets~\cite{Zastrau} and evaporation experiments, e.g. at the Sandia Z-machine~\cite{benage,karasiev_importance,low_density1,low_density2}.
On the other hand, these conditions are highly interesting from a theoretical point of view, as electronic exchange--correlation effects are even more important due to the increased coupling strength~\cite{review,groth_prl,kushal}.

First and foremost, we find that the nonlinear behaviour of the density response appears for significantly smaller perturbation amplitudes as compared to $r_s=2$, which is due to the different energy scales in the system~\cite{energy_note}. For example, for $A=0.1$ the actual response (blue symbols) is suppressed by around $30\%$, whereas hardly any effect would be noticed at the higher density in this case. Overall, both $\chi(q)$ and $\chi_3(q)$ exhibit a similar structure as for $r_s=2$, but are somewhat more symmetric around the maximum at $q\approx2q_\textnormal{F}$. Moreover, $\chi_3(q)$ nearly vanishes for the smallest depicted $q$-value (the leftmost green cross, corresponding to $N=34$) and we find a value more than two orders of magnitude smaller than for $q=2q_\textnormal{F}$. Again, no system-size dependence of $\chi_3(q)$ can be resolved within the given confidence interval even for $N=34$ electrons (crosses).

Another interesting question is how nonlinear effects are influenced by the temperature. To this end, we return to $r_s=2$ for $\theta=1$ (red), $\theta=2$ (green), and $\theta=4$ (blue) in Fig.~\ref{fig:overview}c). With increasing temperature, the linear response function monotonically decreases in magnitude as it is expected, see the bottom half. The same also holds for the cubic response function, where this trend is drastically more pronounced compared to $\chi(q)$. While the maximum in $\chi(q)$ is reduced by a factor of $3$ upon going from $\theta=1$ to $\theta=4$, the cubic response is reduced by a factor of $20$.

\begin{figure}\centering
\includegraphics[width=0.415\textwidth]{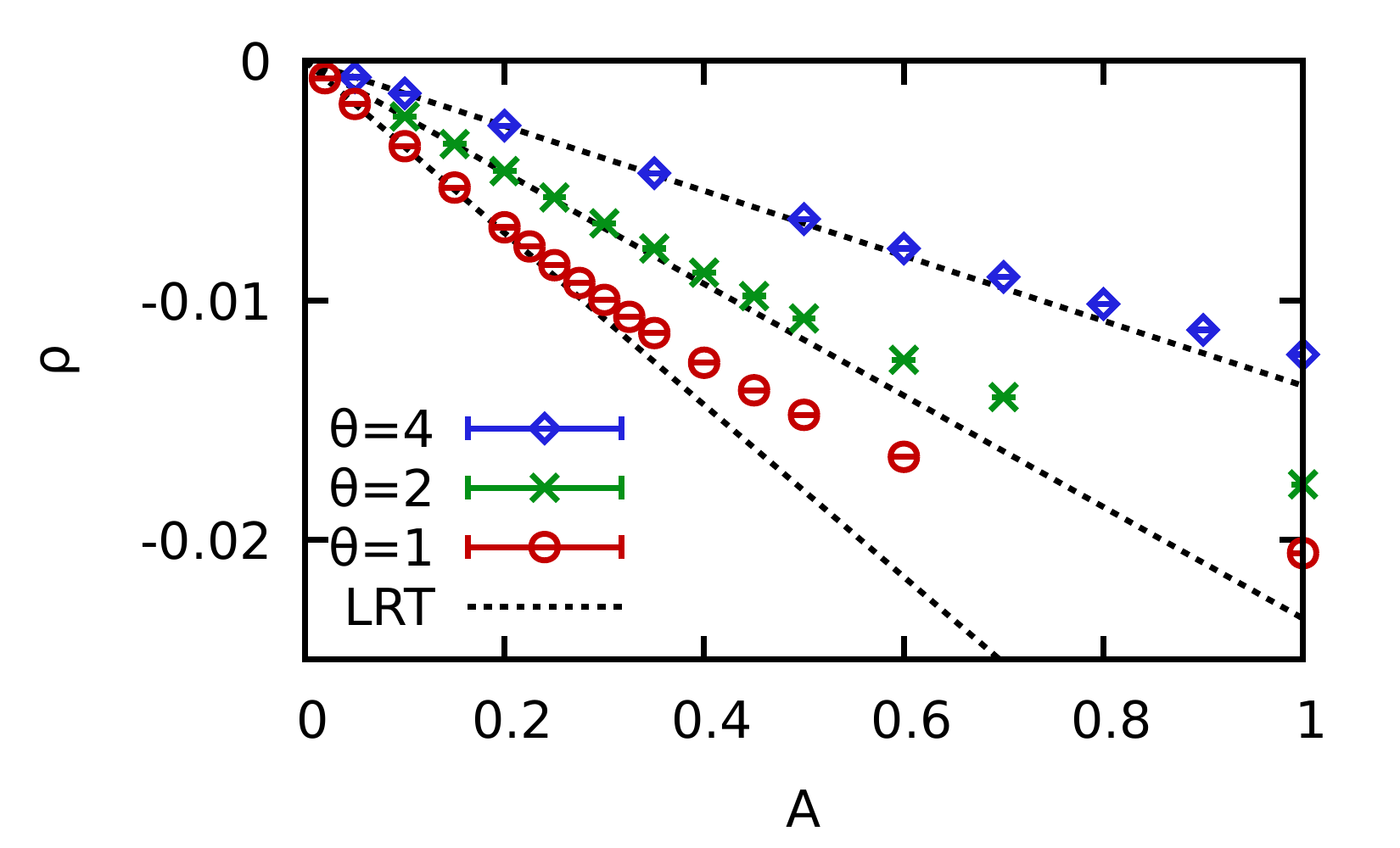}
\caption{\label{fig:THETA}
Density response of the UEG for $N=14$ and $r_s=2$ with $q\approx1.69q_\textnormal{F}$ in dependence of the perturbation amplitude $A$ [cf.~Eq.~(\ref{eq:hamiltonian})]. Shown are PIMC data for the induced density $\rho$ for $\theta=1$ (red circles), $\theta=2$ (green crosses), and $\theta=4$ (blue diamonds), as well as the corresponding predictions from LRT [dotted black, cf.~Eq.~(\ref{eq:static_chi})]. 
}
\end{figure}

This behaviour is further illustrated in Fig.~\ref{fig:THETA}, where we show the $A$-dependence of the induced density for the three temperatures at $q\approx1.69q_\textnormal{F}$, i.e., around the maximum of the density response.
The different symbols correspond to our PIMC data, and the dotted lines to the prediction from LRT, i.e., Eq.~(\ref{eq:static_chi}). There are two dominant trends: 1) the actual density response is smaller for large $\theta$ and 2) LRT remains accurate for larger $A$.


Let us conclude this investigation by briefly touching upon the impact of our findings on state-of-the-art WDM experiments. 
A typical free electron laser with a frequency corresponding to a photon energy of {$8$ keV} and an intensity of {$I\sim10^{17-19}$}W/cm$^2$ corresponds to an approximate perturbation amplitude on the order of $A\sim10^{-5}-10^{-3}$ (see the Supplemental Material~\cite{supplement} for details), which falls safely into the LRT regime even for low densities. On the other hand, intensities of up to {$I=10^{22}$}W/cm$^2$ have been reported recently by employing the novel seeding technique \cite{Fletcher2015}, which results in $A\sim 2$ and clearly violates the boundaries of LRT for both for $r_s=2$ and $r_s=6$. Even current VUV lasers like Flash are capable to reach the nonlinear regime \cite{Zastrau,supplement}.
Another application of our findings concerns the experimental probing of the low-frequency response of WDM using THz lasers \cite{Ofori_Okai_2018}. For example, the recently reported setup with an intensity of $600$kV/cm at around $1$ THz leads to a perturbation amplitude of $A=0.29$ Ha, such that a thorough theoretical interpretation of a corresponding scattering signal would most likely require to take into account nonlinear effects.

\textbf{Summary.} We have carried out extensive \textit{ab initio} PIMC simulations of the harmonically perturbed electron gas. This has allowed us to 1) unambiguously characterize the validity range of LRT and 2) to obtain the first results for the cubic response function $\chi_3(q)$ of the warm dense electron gas, including all exchange--correlation effects. Firstly, we have found that including $\chi_3(q)$ significantly improves the accuracy of the density response function for larger perturbation amplitudes. Moreover, nonlinear effects are particularly important for intermediate wave numbers $q\sim q_\textnormal{F}$, whereas $\chi_3(q)$ vanishes both in the small- and large-$q$ regimes. Regarding physical parameters, we have found that nonlinear effects become more important at lower densities due to the intrinsic energy scale of the system.
This makes materials of relatively low density a highly interesting laboratory to study the interplay of nonlinearity with electronic exchange--correlation effects, and a challenging benchmark for theory.

In addition, we have found that nonlinear effects are severely affected by the electronic temperature and vanish upon increasing $\theta$. While our current simulations are limited to temperatures down to the Fermi temperature ($\theta=1$), this is a strong indication that nonlinear effects might be even more important for lower temperatures $\theta=0.1...0.5$, where many WDM experiments are located.

Our findings are particularly relevant for state-of-the-art WDM experiments with intense free electron lasers in the x-ray or VUV regime, and for low-frequency probing in the THz regime, where the diagnostics methods rely on theory input for the response functions \cite{Fletcher2015,Zastrau,Ofori_Okai_2018}.
Finally, our results will also be important for nonlinear optical diagnostics such as Raman or four-wave mixing spectroscopy, e.g.~\cite{bloembergen_rmp_82,mukamel_86, axt_rmp_98}, or THz streaking \cite{schuette_prl_12} that could provide additional information on correlation effects in warm dense matter.

All PIMC data are available online~\cite{supplement} and can be used to benchmark approximate theories like DFT. Moreover, our new data are exact within the given confidence interval and thus provide the basis for a more general theory of the electronic density response beyond LRT thus further completing our understanding of the electron gas as a fundamental model system~\cite{review,status}.


\section*{Acknowledgments}
We acknowledge stimulating discussions with Richard Pausch and Dominik Kraus, and helpful comments by Michael Bussmann.
This work was partly funded by the Center of Advanced Systems Understanding (CASUS) which is financed by Germany's Federal Ministry of Education and Research (BMBF) and by the Saxon Ministry for Science and Art (SMWK) with tax funds on the basis of the budget approved by the Saxon State Parliament, and by the Deutsche Forschungsgemeinschaft (DFG) via project BO1366/13.
The PIMC calculations were carried out at the Norddeutscher Verbund f\"ur Hoch- und H\"ochstleistungsrechnen (HLRN) under grant shp00015, on a Bull Cluster at the Center for Information Services and High Performace Computing (ZIH) at Technische Universit\"at Dresden,
on the clusters \emph{hypnos} and \emph{hemera} at Helmholtz-Zentrum Dresden-Rossendorf (HZDR), and at the computing center (Rechenzentrum) of Kiel university.

\bibliography{bibliography}{}

\begin{thebibliography}{92}%
\makeatletter
\providecommand \@ifxundefined [1]{%
 \@ifx{#1\undefined}
}%
\providecommand \@ifnum [1]{%
 \ifnum #1\expandafter \@firstoftwo
 \else \expandafter \@secondoftwo
 \fi
}%
\providecommand \@ifx [1]{%
 \ifx #1\expandafter \@firstoftwo
 \else \expandafter \@secondoftwo
 \fi
}%
\providecommand \natexlab [1]{#1}%
\providecommand \enquote  [1]{``#1''}%
\providecommand \bibnamefont  [1]{#1}%
\providecommand \bibfnamefont [1]{#1}%
\providecommand \citenamefont [1]{#1}%
\providecommand \href@noop [0]{\@secondoftwo}%
\providecommand \href [0]{\begingroup \@sanitize@url \@href}%
\providecommand \@href[1]{\@@startlink{#1}\@@href}%
\providecommand \@@href[1]{\endgroup#1\@@endlink}%
\providecommand \@sanitize@url [0]{\catcode `\\12\catcode `\$12\catcode
  `\&12\catcode `\#12\catcode `\^12\catcode `\_12\catcode `\%12\relax}%
\providecommand \@@startlink[1]{}%
\providecommand \@@endlink[0]{}%
\providecommand \url  [0]{\begingroup\@sanitize@url \@url }%
\providecommand \@url [1]{\endgroup\@href {#1}{\urlprefix }}%
\providecommand \urlprefix  [0]{URL }%
\providecommand \Eprint [0]{\href }%
\providecommand \doibase [0]{http://dx.doi.org/}%
\providecommand \selectlanguage [0]{\@gobble}%
\providecommand \bibinfo  [0]{\@secondoftwo}%
\providecommand \bibfield  [0]{\@secondoftwo}%
\providecommand \translation [1]{[#1]}%
\providecommand \BibitemOpen [0]{}%
\providecommand \bibitemStop [0]{}%
\providecommand \bibitemNoStop [0]{.\EOS\space}%
\providecommand \EOS [0]{\spacefactor3000\relax}%
\providecommand \BibitemShut  [1]{\csname bibitem#1\endcsname}%
\let\auto@bib@innerbib\@empty
\bibitem [{\citenamefont {Militzer}\ \emph {et~al.}(2008)\citenamefont
  {Militzer}, \citenamefont {Hubbard}, \citenamefont {Vorberger}, \citenamefont
  {Tamblyn},\ and\ \citenamefont {Bonev}}]{Militzer_2008}%
  \BibitemOpen
  \bibfield  {author} {\bibinfo {author} {\bibfnamefont {B.}~\bibnamefont
  {Militzer}}, \bibinfo {author} {\bibfnamefont {W.~B.}\ \bibnamefont
  {Hubbard}}, \bibinfo {author} {\bibfnamefont {J.}~\bibnamefont {Vorberger}},
  \bibinfo {author} {\bibfnamefont {I.}~\bibnamefont {Tamblyn}}, \ and\
  \bibinfo {author} {\bibfnamefont {S.~A.}\ \bibnamefont {Bonev}},\ }\bibfield
  {title} {\enquote {\bibinfo {title} {A massive core in jupiter predicted from
  first-principles simulations},}\ }\href {\doibase 10.1086/594364} {\bibfield
  {journal} {\bibinfo  {journal} {The Astrophysical Journal}\ }\textbf
  {\bibinfo {volume} {688}},\ \bibinfo {pages} {L45--L48} (\bibinfo {year}
  {2008})}\BibitemShut {NoStop}%
\bibitem [{\citenamefont {Guillot}\ \emph {et~al.}(2018)\citenamefont
  {Guillot}, \citenamefont {Miguel}, \citenamefont {Militzer}, \citenamefont
  {Hubbard}, \citenamefont {Kaspi}, \citenamefont {Galanti}, \citenamefont
  {Cao}, \citenamefont {Helled}, \citenamefont {Wahl}, \citenamefont {Iess},
  \citenamefont {Folkner}, \citenamefont {Stevenson}, \citenamefont {Lunine},
  \citenamefont {Reese}, \citenamefont {Biekman}, \citenamefont {Parisi},
  \citenamefont {Durante}, \citenamefont {Connerney}, \citenamefont {Levin},\
  and\ \citenamefont {Bolton}}]{Guillot2018}%
  \BibitemOpen
  \bibfield  {author} {\bibinfo {author} {\bibfnamefont {T.}~\bibnamefont
  {Guillot}}, \bibinfo {author} {\bibfnamefont {Y.}~\bibnamefont {Miguel}},
  \bibinfo {author} {\bibfnamefont {B.}~\bibnamefont {Militzer}}, \bibinfo
  {author} {\bibfnamefont {W.~B.}\ \bibnamefont {Hubbard}}, \bibinfo {author}
  {\bibfnamefont {Y.}~\bibnamefont {Kaspi}}, \bibinfo {author} {\bibfnamefont
  {E.}~\bibnamefont {Galanti}}, \bibinfo {author} {\bibfnamefont
  {H.}~\bibnamefont {Cao}}, \bibinfo {author} {\bibfnamefont {R.}~\bibnamefont
  {Helled}}, \bibinfo {author} {\bibfnamefont {S.~M.}\ \bibnamefont {Wahl}},
  \bibinfo {author} {\bibfnamefont {L.}~\bibnamefont {Iess}}, \bibinfo {author}
  {\bibfnamefont {W.~M.}\ \bibnamefont {Folkner}}, \bibinfo {author}
  {\bibfnamefont {D.~J.}\ \bibnamefont {Stevenson}}, \bibinfo {author}
  {\bibfnamefont {J.~I.}\ \bibnamefont {Lunine}}, \bibinfo {author}
  {\bibfnamefont {D.~R.}\ \bibnamefont {Reese}}, \bibinfo {author}
  {\bibfnamefont {A.}~\bibnamefont {Biekman}}, \bibinfo {author} {\bibfnamefont
  {M.}~\bibnamefont {Parisi}}, \bibinfo {author} {\bibfnamefont
  {D.}~\bibnamefont {Durante}}, \bibinfo {author} {\bibfnamefont {J.~E.~P.}\
  \bibnamefont {Connerney}}, \bibinfo {author} {\bibfnamefont {S.~M.}\
  \bibnamefont {Levin}}, \ and\ \bibinfo {author} {\bibfnamefont {S.~J.}\
  \bibnamefont {Bolton}},\ }\bibfield  {title} {\enquote {\bibinfo {title} {A
  suppression of differential rotation in jupiter's deep interior},}\ }\href
  {\doibase 10.1038/nature25775} {\bibfield  {journal} {\bibinfo  {journal}
  {Nature}\ }\textbf {\bibinfo {volume} {555}},\ \bibinfo {pages} {227--230}
  (\bibinfo {year} {2018})}\BibitemShut {NoStop}%
\bibitem [{\citenamefont {Saumon}\ \emph {et~al.}(1992)\citenamefont {Saumon},
  \citenamefont {Hubbard}, \citenamefont {Chabrier},\ and\ \citenamefont {van
  Horn}}]{saumon1}%
  \BibitemOpen
  \bibfield  {author} {\bibinfo {author} {\bibfnamefont {D.}~\bibnamefont
  {Saumon}}, \bibinfo {author} {\bibfnamefont {W.~B.}\ \bibnamefont {Hubbard}},
  \bibinfo {author} {\bibfnamefont {G.}~\bibnamefont {Chabrier}}, \ and\
  \bibinfo {author} {\bibfnamefont {H.~M.}\ \bibnamefont {van Horn}},\
  }\bibfield  {title} {\enquote {\bibinfo {title} {The role of the
  molecular-metallic transition of hydrogen in the evolution of jupiter,
  saturn, and brown dwarfs},}\ }\href
  {http://adsabs.harvard.edu/full/1992ApJ...391..827S} {\bibfield  {journal}
  {\bibinfo  {journal} {Astrophys. J}\ }\textbf {\bibinfo {volume} {391}},\
  \bibinfo {pages} {827--831} (\bibinfo {year} {1992})}\BibitemShut {NoStop}%
\bibitem [{\citenamefont {Becker}\ \emph {et~al.}(2014)\citenamefont {Becker},
  \citenamefont {Lorenzen}, \citenamefont {Fortney}, \citenamefont
  {Nettelmann}, \citenamefont {Sch\"ottler},\ and\ \citenamefont
  {Redmer}}]{becker}%
  \BibitemOpen
  \bibfield  {author} {\bibinfo {author} {\bibfnamefont {A.}~\bibnamefont
  {Becker}}, \bibinfo {author} {\bibfnamefont {W.}~\bibnamefont {Lorenzen}},
  \bibinfo {author} {\bibfnamefont {J.~J.}\ \bibnamefont {Fortney}}, \bibinfo
  {author} {\bibfnamefont {N.}~\bibnamefont {Nettelmann}}, \bibinfo {author}
  {\bibfnamefont {M.}~\bibnamefont {Sch\"ottler}}, \ and\ \bibinfo {author}
  {\bibfnamefont {R.}~\bibnamefont {Redmer}},\ }\bibfield  {title} {\enquote
  {\bibinfo {title} {Ab initio equations of state for hydrogen (h-reos.3) and
  helium (he-reos.3) and their implications for the interior of brown
  dwarfs},}\ }\href
  {https://iopscience.iop.org/article/10.1088/0067-0049/215/2/21/meta}
  {\bibfield  {journal} {\bibinfo  {journal} {Astrophys. J. Suppl. Ser}\
  }\textbf {\bibinfo {volume} {215}},\ \bibinfo {pages} {21} (\bibinfo {year}
  {2014})}\BibitemShut {NoStop}%
\bibitem [{\citenamefont {Ernstorfer}\ \emph {et~al.}(2009)\citenamefont
  {Ernstorfer}, \citenamefont {Harb}, \citenamefont {Hebeisen}, \citenamefont
  {Sciaini}, \citenamefont {Dartigalongue},\ and\ \citenamefont
  {Miller}}]{ernstorfer2}%
  \BibitemOpen
  \bibfield  {author} {\bibinfo {author} {\bibfnamefont {R.}~\bibnamefont
  {Ernstorfer}}, \bibinfo {author} {\bibfnamefont {M.}~\bibnamefont {Harb}},
  \bibinfo {author} {\bibfnamefont {C.~T.}\ \bibnamefont {Hebeisen}}, \bibinfo
  {author} {\bibfnamefont {G.}~\bibnamefont {Sciaini}}, \bibinfo {author}
  {\bibfnamefont {T.}~\bibnamefont {Dartigalongue}}, \ and\ \bibinfo {author}
  {\bibfnamefont {R.~J.~D.}\ \bibnamefont {Miller}},\ }\bibfield  {title}
  {\enquote {\bibinfo {title} {The formation of warm dense matter: Experimental
  evidence for electronic bond hardening in gold},}\ }\href
  {http://science.sciencemag.org/content/323/5917/1033} {\bibfield  {journal}
  {\bibinfo  {journal} {Science}\ }\textbf {\bibinfo {volume} {323}},\ \bibinfo
  {pages} {1033} (\bibinfo {year} {2009})}\BibitemShut {NoStop}%
\bibitem [{\citenamefont {Waldecker}\ \emph {et~al.}(2016)\citenamefont
  {Waldecker}, \citenamefont {Bertoni}, \citenamefont {Ernstorfer},\ and\
  \citenamefont {Vorberger}}]{ernstorfer}%
  \BibitemOpen
  \bibfield  {author} {\bibinfo {author} {\bibfnamefont {L.}~\bibnamefont
  {Waldecker}}, \bibinfo {author} {\bibfnamefont {R.}~\bibnamefont {Bertoni}},
  \bibinfo {author} {\bibfnamefont {R.}~\bibnamefont {Ernstorfer}}, \ and\
  \bibinfo {author} {\bibfnamefont {J.}~\bibnamefont {Vorberger}},\ }\bibfield
  {title} {\enquote {\bibinfo {title} {Electron-phonon coupling and energy flow
  in a simple metal beyond the two-temperature approximation},}\ }\href
  {https://journals.aps.org/prx/abstract/10.1103/PhysRevX.6.021003} {\bibfield
  {journal} {\bibinfo  {journal} {Phys. Rev. X}\ }\textbf {\bibinfo {volume}
  {6}},\ \bibinfo {pages} {021003} (\bibinfo {year} {2016})}\BibitemShut
  {NoStop}%
\bibitem [{\citenamefont {Hu}\ \emph {et~al.}(2011)\citenamefont {Hu},
  \citenamefont {Militzer}, \citenamefont {Goncharov},\ and\ \citenamefont
  {Skupsky}}]{hu_ICF}%
  \BibitemOpen
  \bibfield  {author} {\bibinfo {author} {\bibfnamefont {S.~X.}\ \bibnamefont
  {Hu}}, \bibinfo {author} {\bibfnamefont {B.}~\bibnamefont {Militzer}},
  \bibinfo {author} {\bibfnamefont {V.~N.}\ \bibnamefont {Goncharov}}, \ and\
  \bibinfo {author} {\bibfnamefont {S.}~\bibnamefont {Skupsky}},\ }\bibfield
  {title} {\enquote {\bibinfo {title} {First-principles equation-of-state table
  of deuterium for inertial confinement fusion applications},}\ }\href
  {https://journals.aps.org/prb/abstract/10.1103/PhysRevB.84.224109} {\bibfield
   {journal} {\bibinfo  {journal} {Phys. Rev. B}\ }\textbf {\bibinfo {volume}
  {84}},\ \bibinfo {pages} {224109} (\bibinfo {year} {2011})}\BibitemShut
  {NoStop}%
\bibitem [{\citenamefont {Fortov}(2009)}]{fortov_review}%
  \BibitemOpen
  \bibfield  {author} {\bibinfo {author} {\bibfnamefont {V.~E.}\ \bibnamefont
  {Fortov}},\ }\bibfield  {title} {\enquote {\bibinfo {title} {Extreme states
  of matter on earth and in space},}\ }\href
  {https://www.turpion.org/php/paper.phtml?journal_id=pu&paper_id=6821}
  {\bibfield  {journal} {\bibinfo  {journal} {Phys.-Usp}\ }\textbf {\bibinfo
  {volume} {52}},\ \bibinfo {pages} {615--647} (\bibinfo {year}
  {2009})}\BibitemShut {NoStop}%
\bibitem [{\citenamefont {Glenzer}\ and\ \citenamefont
  {Redmer}(2009)}]{siegfried_review}%
  \BibitemOpen
  \bibfield  {author} {\bibinfo {author} {\bibfnamefont {S.~H.}\ \bibnamefont
  {Glenzer}}\ and\ \bibinfo {author} {\bibfnamefont {R.}~\bibnamefont
  {Redmer}},\ }\bibfield  {title} {\enquote {\bibinfo {title} {X-ray thomson
  scattering in high energy density plasmas},}\ }\href
  {https://journals.aps.org/rmp/abstract/10.1103/RevModPhys.81.1625} {\bibfield
   {journal} {\bibinfo  {journal} {Rev. Mod. Phys}\ }\textbf {\bibinfo {volume}
  {81}},\ \bibinfo {pages} {1625} (\bibinfo {year} {2009})}\BibitemShut
  {NoStop}%
\bibitem [{\citenamefont {Falk}(2018)}]{falk_wdm}%
  \BibitemOpen
  \bibfield  {author} {\bibinfo {author} {\bibfnamefont {K.}~\bibnamefont
  {Falk}},\ }\bibfield  {title} {\enquote {\bibinfo {title} {Experimental
  methods for warm dense matter research},}\ }\href
  {https://www.cambridge.org/core/journals/high-power-laser-science-and-engineering/article/experimental-methods-for-warm-dense-matter-research/7205AE1029BEA0061044F84875F1CEDB}
  {\bibfield  {journal} {\bibinfo  {journal} {High Power Laser Sci. Eng}\
  }\textbf {\bibinfo {volume} {6}},\ \bibinfo {pages} {e59} (\bibinfo {year}
  {2018})}\BibitemShut {NoStop}%
\bibitem [{\citenamefont {Moses}\ \emph {et~al.}(2009)\citenamefont {Moses},
  \citenamefont {Boyd}, \citenamefont {Remington}, \citenamefont {Keane},\ and\
  \citenamefont {Al-Ayat}}]{Moses_NIF}%
  \BibitemOpen
  \bibfield  {author} {\bibinfo {author} {\bibfnamefont {E.~I.}\ \bibnamefont
  {Moses}}, \bibinfo {author} {\bibfnamefont {R.~N.}\ \bibnamefont {Boyd}},
  \bibinfo {author} {\bibfnamefont {B.~A.}\ \bibnamefont {Remington}}, \bibinfo
  {author} {\bibfnamefont {C.~J.}\ \bibnamefont {Keane}}, \ and\ \bibinfo
  {author} {\bibfnamefont {R.}~\bibnamefont {Al-Ayat}},\ }\bibfield  {title}
  {\enquote {\bibinfo {title} {The national ignition facility: Ushering in a
  new age for high energy density science},}\ }\href {\doibase
  10.1063/1.3116505} {\bibfield  {journal} {\bibinfo  {journal} {Physics of
  Plasmas}\ }\textbf {\bibinfo {volume} {16}},\ \bibinfo {pages} {041006}
  (\bibinfo {year} {2009})},\ \Eprint
  {http://arxiv.org/abs/https://doi.org/10.1063/1.3116505}
  {https://doi.org/10.1063/1.3116505} \BibitemShut {NoStop}%
\bibitem [{\citenamefont {Bostedt}\ \emph {et~al.}(2016)\citenamefont
  {Bostedt}, \citenamefont {Boutet}, \citenamefont {Fritz}, \citenamefont
  {Huang}, \citenamefont {Lee}, \citenamefont {Lemke}, \citenamefont {Robert},
  \citenamefont {Schlotter}, \citenamefont {Turner},\ and\ \citenamefont
  {Williams}}]{LCLS_2016}%
  \BibitemOpen
  \bibfield  {author} {\bibinfo {author} {\bibfnamefont {Christoph}\
  \bibnamefont {Bostedt}}, \bibinfo {author} {\bibfnamefont {S\'ebastien}\
  \bibnamefont {Boutet}}, \bibinfo {author} {\bibfnamefont {David~M.}\
  \bibnamefont {Fritz}}, \bibinfo {author} {\bibfnamefont {Zhirong}\
  \bibnamefont {Huang}}, \bibinfo {author} {\bibfnamefont {Hae~Ja}\
  \bibnamefont {Lee}}, \bibinfo {author} {\bibfnamefont {Henrik~T.}\
  \bibnamefont {Lemke}}, \bibinfo {author} {\bibfnamefont {Aymeric}\
  \bibnamefont {Robert}}, \bibinfo {author} {\bibfnamefont {William~F.}\
  \bibnamefont {Schlotter}}, \bibinfo {author} {\bibfnamefont {Joshua~J.}\
  \bibnamefont {Turner}}, \ and\ \bibinfo {author} {\bibfnamefont {Garth~J.}\
  \bibnamefont {Williams}},\ }\bibfield  {title} {\enquote {\bibinfo {title}
  {Linac coherent light source: The first five years},}\ }\href {\doibase
  10.1103/RevModPhys.88.015007} {\bibfield  {journal} {\bibinfo  {journal}
  {Rev. Mod. Phys.}\ }\textbf {\bibinfo {volume} {88}},\ \bibinfo {pages}
  {015007} (\bibinfo {year} {2016})}\BibitemShut {NoStop}%
\bibitem [{\citenamefont {Tschentscher}\ \emph {et~al.}(2017)\citenamefont
  {Tschentscher}, \citenamefont {Bressler}, \citenamefont {Grünert},
  \citenamefont {Madsen}, \citenamefont {Mancuso}, \citenamefont {Meyer},
  \citenamefont {Scherz}, \citenamefont {Sinn},\ and\ \citenamefont
  {Zastrau}}]{Tschentscher_2017}%
  \BibitemOpen
  \bibfield  {author} {\bibinfo {author} {\bibfnamefont {Thomas}\ \bibnamefont
  {Tschentscher}}, \bibinfo {author} {\bibfnamefont {Christian}\ \bibnamefont
  {Bressler}}, \bibinfo {author} {\bibfnamefont {Jan}\ \bibnamefont
  {Grünert}}, \bibinfo {author} {\bibfnamefont {Anders}\ \bibnamefont
  {Madsen}}, \bibinfo {author} {\bibfnamefont {Adrian~P.}\ \bibnamefont
  {Mancuso}}, \bibinfo {author} {\bibfnamefont {Michael}\ \bibnamefont
  {Meyer}}, \bibinfo {author} {\bibfnamefont {Andreas}\ \bibnamefont {Scherz}},
  \bibinfo {author} {\bibfnamefont {Harald}\ \bibnamefont {Sinn}}, \ and\
  \bibinfo {author} {\bibfnamefont {Ulf}\ \bibnamefont {Zastrau}},\ }\bibfield
  {title} {\enquote {\bibinfo {title} {Photon beam transport and scientific
  instruments at the european xfel},}\ }\href {\doibase 10.3390/app7060592}
  {\bibfield  {journal} {\bibinfo  {journal} {Applied Sciences}\ }\textbf
  {\bibinfo {volume} {7}} (\bibinfo {year} {2017}),\
  10.3390/app7060592}\BibitemShut {NoStop}%
\bibitem [{\citenamefont {Bonitz}\ \emph {et~al.}()\citenamefont {Bonitz},
  \citenamefont {Dornheim}, \citenamefont {Moldabekov}, \citenamefont {Zhang},
  \citenamefont {Hamann}, \citenamefont {Filinov}, \citenamefont
  {Ramakrishna},\ and\ \citenamefont {Vorberger}}]{new_POP}%
  \BibitemOpen
  \bibfield  {author} {\bibinfo {author} {\bibfnamefont {M.}~\bibnamefont
  {Bonitz}}, \bibinfo {author} {\bibfnamefont {T.}~\bibnamefont {Dornheim}},
  \bibinfo {author} {\bibfnamefont {Zh.A.}\ \bibnamefont {Moldabekov}},
  \bibinfo {author} {\bibfnamefont {S.}~\bibnamefont {Zhang}}, \bibinfo
  {author} {\bibfnamefont {P.}~\bibnamefont {Hamann}}, \bibinfo {author}
  {\bibfnamefont {A.}~\bibnamefont {Filinov}}, \bibinfo {author} {\bibfnamefont
  {K.}~\bibnamefont {Ramakrishna}}, \ and\ \bibinfo {author} {\bibfnamefont
  {J.}~\bibnamefont {Vorberger}},\ }\href {https://arxiv.org/abs/1912.09884}
  {}\ (\bibinfo  {publisher} {Ab initio simulation of warm dense matter})\
  \Eprint {http://arxiv.org/abs/1912.09884} {arXiv:1912.09884} \BibitemShut
  {NoStop}%
\bibitem [{\citenamefont {Graziani}\ \emph {et~al.}(2014)\citenamefont
  {Graziani}, \citenamefont {Desjarlais}, \citenamefont {Redmer},\ and\
  \citenamefont {Trickey}}]{wdm_book}%
  \BibitemOpen
  \bibinfo {editor} {\bibfnamefont {F.}~\bibnamefont {Graziani}}, \bibinfo
  {editor} {\bibfnamefont {M.~P.}\ \bibnamefont {Desjarlais}}, \bibinfo
  {editor} {\bibfnamefont {R.}~\bibnamefont {Redmer}}, \ and\ \bibinfo {editor}
  {\bibfnamefont {S.~B.}\ \bibnamefont {Trickey}},\ eds.,\ \href@noop {} {\emph
  {\bibinfo {title} {Frontiers and Challenges in Warm Dense Matter}}}\
  (\bibinfo  {publisher} {Springer},\ \bibinfo {address} {International
  Publishing},\ \bibinfo {year} {2014})\BibitemShut {NoStop}%
\bibitem [{\citenamefont {Kraus}\ \emph {et~al.}(2019)\citenamefont {Kraus},
  \citenamefont {Bachmann}, \citenamefont {Barbrel}, \citenamefont {Falcone},
  \citenamefont {Fletcher}, \citenamefont {Frydrych}, \citenamefont {Gamboa},
  \citenamefont {Gauthier}, \citenamefont {Gericke}, \citenamefont {Glenzer},
  \citenamefont {G\"ode}, \citenamefont {Granados}, \citenamefont {Hartley},
  \citenamefont {Helfrich}, \citenamefont {Lee}, \citenamefont {Nagler},
  \citenamefont {Ravasio}, \citenamefont {Schumaker}, \citenamefont
  {Vorberger},\ and\ \citenamefont {D\"oppner}}]{kraus_xrts}%
  \BibitemOpen
  \bibfield  {author} {\bibinfo {author} {\bibfnamefont {D.}~\bibnamefont
  {Kraus}}, \bibinfo {author} {\bibfnamefont {B.}~\bibnamefont {Bachmann}},
  \bibinfo {author} {\bibfnamefont {B.}~\bibnamefont {Barbrel}}, \bibinfo
  {author} {\bibfnamefont {R.~W.}\ \bibnamefont {Falcone}}, \bibinfo {author}
  {\bibfnamefont {L.~B.}\ \bibnamefont {Fletcher}}, \bibinfo {author}
  {\bibfnamefont {S.}~\bibnamefont {Frydrych}}, \bibinfo {author}
  {\bibfnamefont {E.~J.}\ \bibnamefont {Gamboa}}, \bibinfo {author}
  {\bibfnamefont {M.}~\bibnamefont {Gauthier}}, \bibinfo {author}
  {\bibfnamefont {D.~O.}\ \bibnamefont {Gericke}}, \bibinfo {author}
  {\bibfnamefont {S.~H.}\ \bibnamefont {Glenzer}}, \bibinfo {author}
  {\bibfnamefont {S.}~\bibnamefont {G\"ode}}, \bibinfo {author} {\bibfnamefont
  {E.}~\bibnamefont {Granados}}, \bibinfo {author} {\bibfnamefont {N.~J.}\
  \bibnamefont {Hartley}}, \bibinfo {author} {\bibfnamefont {J.}~\bibnamefont
  {Helfrich}}, \bibinfo {author} {\bibfnamefont {H.~J.}\ \bibnamefont {Lee}},
  \bibinfo {author} {\bibfnamefont {B.}~\bibnamefont {Nagler}}, \bibinfo
  {author} {\bibfnamefont {A.}~\bibnamefont {Ravasio}}, \bibinfo {author}
  {\bibfnamefont {W.}~\bibnamefont {Schumaker}}, \bibinfo {author}
  {\bibfnamefont {J.}~\bibnamefont {Vorberger}}, \ and\ \bibinfo {author}
  {\bibfnamefont {T.}~\bibnamefont {D\"oppner}},\ }\bibfield  {title} {\enquote
  {\bibinfo {title} {Characterizing the ionization potential depression in
  dense carbon plasmas with high-precision spectrally resolved x-ray
  scattering},}\ }\href
  {https://iopscience.iop.org/article/10.1088/1361-6587/aadd6c/meta} {\bibfield
   {journal} {\bibinfo  {journal} {Plasma Phys. Control Fusion}\ }\textbf
  {\bibinfo {volume} {61}},\ \bibinfo {pages} {014015} (\bibinfo {year}
  {2019})}\BibitemShut {NoStop}%
\bibitem [{\citenamefont {Cayzac}\ \emph {et~al.}(2017)\citenamefont {Cayzac},
  \citenamefont {Frank}, \citenamefont {Ortner}, \citenamefont {Bagnoud},
  \citenamefont {Basko}, \citenamefont {Bedacht}, \citenamefont {Bl{\"a}ser},
  \citenamefont {Blazevic}, \citenamefont {Busold}, \citenamefont {Deppert},
  \citenamefont {Ding}, \citenamefont {Ehret}, \citenamefont {Fiala},
  \citenamefont {Frydrych}, \citenamefont {Gericke}, \citenamefont {Hallo},
  \citenamefont {Helfrich}, \citenamefont {Jahn}, \citenamefont {Kjartansson},
  \citenamefont {Knetsch}, \citenamefont {Kraus}, \citenamefont {Malka},
  \citenamefont {Neumann}, \citenamefont {P{\'e}pitone}, \citenamefont
  {Pepler}, \citenamefont {Sander}, \citenamefont {Schaumann}, \citenamefont
  {Schlegel}, \citenamefont {Schroeter}, \citenamefont {Schumacher},
  \citenamefont {Seibert}, \citenamefont {Tauschwitz}, \citenamefont
  {Vorberger}, \citenamefont {Wagner}, \citenamefont {Weih}, \citenamefont
  {Zobus},\ and\ \citenamefont {Roth}}]{Cayzac2017}%
  \BibitemOpen
  \bibfield  {author} {\bibinfo {author} {\bibfnamefont {W.}~\bibnamefont
  {Cayzac}}, \bibinfo {author} {\bibfnamefont {A.}~\bibnamefont {Frank}},
  \bibinfo {author} {\bibfnamefont {A.}~\bibnamefont {Ortner}}, \bibinfo
  {author} {\bibfnamefont {V.}~\bibnamefont {Bagnoud}}, \bibinfo {author}
  {\bibfnamefont {M.~M.}\ \bibnamefont {Basko}}, \bibinfo {author}
  {\bibfnamefont {S.}~\bibnamefont {Bedacht}}, \bibinfo {author} {\bibfnamefont
  {C.}~\bibnamefont {Bl{\"a}ser}}, \bibinfo {author} {\bibfnamefont
  {A.}~\bibnamefont {Blazevic}}, \bibinfo {author} {\bibfnamefont
  {S.}~\bibnamefont {Busold}}, \bibinfo {author} {\bibfnamefont
  {O.}~\bibnamefont {Deppert}}, \bibinfo {author} {\bibfnamefont
  {J.}~\bibnamefont {Ding}}, \bibinfo {author} {\bibfnamefont {M.}~\bibnamefont
  {Ehret}}, \bibinfo {author} {\bibfnamefont {P.}~\bibnamefont {Fiala}},
  \bibinfo {author} {\bibfnamefont {S.}~\bibnamefont {Frydrych}}, \bibinfo
  {author} {\bibfnamefont {D.~O.}\ \bibnamefont {Gericke}}, \bibinfo {author}
  {\bibfnamefont {L.}~\bibnamefont {Hallo}}, \bibinfo {author} {\bibfnamefont
  {J.}~\bibnamefont {Helfrich}}, \bibinfo {author} {\bibfnamefont
  {D.}~\bibnamefont {Jahn}}, \bibinfo {author} {\bibfnamefont {E.}~\bibnamefont
  {Kjartansson}}, \bibinfo {author} {\bibfnamefont {A.}~\bibnamefont
  {Knetsch}}, \bibinfo {author} {\bibfnamefont {D.}~\bibnamefont {Kraus}},
  \bibinfo {author} {\bibfnamefont {G.}~\bibnamefont {Malka}}, \bibinfo
  {author} {\bibfnamefont {N.~W.}\ \bibnamefont {Neumann}}, \bibinfo {author}
  {\bibfnamefont {K.}~\bibnamefont {P{\'e}pitone}}, \bibinfo {author}
  {\bibfnamefont {D.}~\bibnamefont {Pepler}}, \bibinfo {author} {\bibfnamefont
  {S.}~\bibnamefont {Sander}}, \bibinfo {author} {\bibfnamefont
  {G.}~\bibnamefont {Schaumann}}, \bibinfo {author} {\bibfnamefont
  {T.}~\bibnamefont {Schlegel}}, \bibinfo {author} {\bibfnamefont
  {N.}~\bibnamefont {Schroeter}}, \bibinfo {author} {\bibfnamefont
  {D.}~\bibnamefont {Schumacher}}, \bibinfo {author} {\bibfnamefont
  {M.}~\bibnamefont {Seibert}}, \bibinfo {author} {\bibfnamefont
  {An}~\bibnamefont {Tauschwitz}}, \bibinfo {author} {\bibfnamefont
  {J.}~\bibnamefont {Vorberger}}, \bibinfo {author} {\bibfnamefont
  {F.}~\bibnamefont {Wagner}}, \bibinfo {author} {\bibfnamefont
  {S.}~\bibnamefont {Weih}}, \bibinfo {author} {\bibfnamefont {Y.}~\bibnamefont
  {Zobus}}, \ and\ \bibinfo {author} {\bibfnamefont {M.}~\bibnamefont {Roth}},\
  }\bibfield  {title} {\enquote {\bibinfo {title} {Experimental discrimination
  of ion stopping models near the bragg peak in highly ionized matter},}\
  }\href {\doibase 10.1038/ncomms15693} {\bibfield  {journal} {\bibinfo
  {journal} {Nature Communications}\ }\textbf {\bibinfo {volume} {8}},\
  \bibinfo {pages} {15693} (\bibinfo {year} {2017})}\BibitemShut {NoStop}%
\bibitem [{\citenamefont {Senatore}\ \emph {et~al.}(1996)\citenamefont
  {Senatore}, \citenamefont {Moroni},\ and\ \citenamefont
  {Ceperley}}]{ceperley_potential}%
  \BibitemOpen
  \bibfield  {author} {\bibinfo {author} {\bibfnamefont {G.}~\bibnamefont
  {Senatore}}, \bibinfo {author} {\bibfnamefont {S.}~\bibnamefont {Moroni}}, \
  and\ \bibinfo {author} {\bibfnamefont {D.~M.}\ \bibnamefont {Ceperley}},\
  }\bibfield  {title} {\enquote {\bibinfo {title} {Local field factor and
  effective potentials in liquid metals},}\ }\href
  {https://www.sciencedirect.com/science/article/pii/S002230939600316X}
  {\bibfield  {journal} {\bibinfo  {journal} {J. Non-Cryst. Sol}\ }\textbf
  {\bibinfo {volume} {205-207}},\ \bibinfo {pages} {851--854} (\bibinfo {year}
  {1996})}\BibitemShut {NoStop}%
\bibitem [{\citenamefont {Moldabekov}\ \emph {et~al.}(2018)\citenamefont
  {Moldabekov}, \citenamefont {Groth}, \citenamefont {Dornheim}, \citenamefont
  {K\"ahlert}, \citenamefont {Bonitz},\ and\ \citenamefont
  {Ramazanov}}]{zhandos1}%
  \BibitemOpen
  \bibfield  {author} {\bibinfo {author} {\bibfnamefont {Zh.A.}\ \bibnamefont
  {Moldabekov}}, \bibinfo {author} {\bibfnamefont {S.}~\bibnamefont {Groth}},
  \bibinfo {author} {\bibfnamefont {T.}~\bibnamefont {Dornheim}}, \bibinfo
  {author} {\bibfnamefont {H.}~\bibnamefont {K\"ahlert}}, \bibinfo {author}
  {\bibfnamefont {M.}~\bibnamefont {Bonitz}}, \ and\ \bibinfo {author}
  {\bibfnamefont {T.~S.}\ \bibnamefont {Ramazanov}},\ }\bibfield  {title}
  {\enquote {\bibinfo {title} {Structural characteristics of strongly coupled
  ions in a dense quantum plasma},}\ }\href
  {https://journals.aps.org/pre/abstract/10.1103/PhysRevE.98.023207} {\bibfield
   {journal} {\bibinfo  {journal} {Phys. Rev. E}\ }\textbf {\bibinfo {volume}
  {98}},\ \bibinfo {pages} {023207} (\bibinfo {year} {2018})}\BibitemShut
  {NoStop}%
\bibitem [{\citenamefont {Moldabekov}\ \emph {et~al.}(2019)\citenamefont
  {Moldabekov}, \citenamefont {K\"ahlert}, \citenamefont {Dornheim},
  \citenamefont {Groth}, \citenamefont {Bonitz},\ and\ \citenamefont
  {Ramazanov}}]{zhandos2}%
  \BibitemOpen
  \bibfield  {author} {\bibinfo {author} {\bibfnamefont {Zh.A.}\ \bibnamefont
  {Moldabekov}}, \bibinfo {author} {\bibfnamefont {H.}~\bibnamefont
  {K\"ahlert}}, \bibinfo {author} {\bibfnamefont {T.}~\bibnamefont {Dornheim}},
  \bibinfo {author} {\bibfnamefont {S.}~\bibnamefont {Groth}}, \bibinfo
  {author} {\bibfnamefont {M.}~\bibnamefont {Bonitz}}, \ and\ \bibinfo {author}
  {\bibfnamefont {T.~S.}\ \bibnamefont {Ramazanov}},\ }\bibfield  {title}
  {\enquote {\bibinfo {title} {Dynamical structure factor of strongly coupled
  ions in a dense quantum plasma},}\ }\href
  {https://journals.aps.org/pre/abstract/10.1103/PhysRevE.99.053203} {\bibfield
   {journal} {\bibinfo  {journal} {Phys. Rev. E}\ }\textbf {\bibinfo {volume}
  {99}},\ \bibinfo {pages} {053203} (\bibinfo {year} {2019})}\BibitemShut
  {NoStop}%
\bibitem [{\citenamefont {Baczewski}\ \emph {et~al.}(2016)\citenamefont
  {Baczewski}, \citenamefont {Shulenburger}, \citenamefont {Desjarlais},
  \citenamefont {Hansen},\ and\ \citenamefont {Magyar}}]{dynamic2}%
  \BibitemOpen
  \bibfield  {author} {\bibinfo {author} {\bibfnamefont {A.~D.}\ \bibnamefont
  {Baczewski}}, \bibinfo {author} {\bibfnamefont {L.}~\bibnamefont
  {Shulenburger}}, \bibinfo {author} {\bibfnamefont {M.~P.}\ \bibnamefont
  {Desjarlais}}, \bibinfo {author} {\bibfnamefont {S.~B.}\ \bibnamefont
  {Hansen}}, \ and\ \bibinfo {author} {\bibfnamefont {R.~J.}\ \bibnamefont
  {Magyar}},\ }\bibfield  {title} {\enquote {\bibinfo {title} {X-ray thomson
  scattering in warm dense matter without the chihara decomposition},}\ }\href
  {https://journals.aps.org/prl/abstract/10.1103/PhysRevLett.116.115004}
  {\bibfield  {journal} {\bibinfo  {journal} {Phys. Rev. Lett}\ }\textbf
  {\bibinfo {volume} {116}},\ \bibinfo {pages} {115004} (\bibinfo {year}
  {2016})}\BibitemShut {NoStop}%
\bibitem [{\citenamefont {Pribram-Jones}\ \emph {et~al.}(2016)\citenamefont
  {Pribram-Jones}, \citenamefont {Grabowski},\ and\ \citenamefont
  {Burke}}]{pribram}%
  \BibitemOpen
  \bibfield  {author} {\bibinfo {author} {\bibfnamefont {A.}~\bibnamefont
  {Pribram-Jones}}, \bibinfo {author} {\bibfnamefont {P.~E.}\ \bibnamefont
  {Grabowski}}, \ and\ \bibinfo {author} {\bibfnamefont {K.}~\bibnamefont
  {Burke}},\ }\bibfield  {title} {\enquote {\bibinfo {title} {Thermal density
  functional theory: Time-dependent linear response and approximate functionals
  from the fluctuation-dissipation theorem},}\ }\href
  {https://journals.aps.org/prl/abstract/10.1103/PhysRevLett.116.233001}
  {\bibfield  {journal} {\bibinfo  {journal} {Phys. Rev. Lett}\ }\textbf
  {\bibinfo {volume} {116}},\ \bibinfo {pages} {233001} (\bibinfo {year}
  {2016})}\BibitemShut {NoStop}%
\bibitem [{\citenamefont {Vorberger}\ \emph {et~al.}(2010)\citenamefont
  {Vorberger}, \citenamefont {Gericke}, \citenamefont {Bornath},\ and\
  \citenamefont {Schlanges}}]{transfer1}%
  \BibitemOpen
  \bibfield  {author} {\bibinfo {author} {\bibfnamefont {J.}~\bibnamefont
  {Vorberger}}, \bibinfo {author} {\bibfnamefont {D.~O.}\ \bibnamefont
  {Gericke}}, \bibinfo {author} {\bibfnamefont {Th.}\ \bibnamefont {Bornath}},
  \ and\ \bibinfo {author} {\bibfnamefont {M.}~\bibnamefont {Schlanges}},\
  }\bibfield  {title} {\enquote {\bibinfo {title} {Energy relaxation in dense,
  strongly coupled two-temperature plasmas},}\ }\href
  {https://journals.aps.org/pre/abstract/10.1103/PhysRevE.96.023203} {\bibfield
   {journal} {\bibinfo  {journal} {Phys. Rev. E}\ }\textbf {\bibinfo {volume}
  {81}},\ \bibinfo {pages} {046404} (\bibinfo {year} {2010})}\BibitemShut
  {NoStop}%
\bibitem [{\citenamefont {Benedict}\ \emph {et~al.}(2017)\citenamefont
  {Benedict}, \citenamefont {Surh}, \citenamefont {Stanton}, \citenamefont
  {Scullard}, \citenamefont {Correa}, \citenamefont {Castor}, \citenamefont
  {Graziani}, \citenamefont {Collins}, \citenamefont {Certík}, \citenamefont
  {Kress},\ and\ \citenamefont {Murillo}}]{transfer2}%
  \BibitemOpen
  \bibfield  {author} {\bibinfo {author} {\bibfnamefont {L.~X.}\ \bibnamefont
  {Benedict}}, \bibinfo {author} {\bibfnamefont {M.~P.}\ \bibnamefont {Surh}},
  \bibinfo {author} {\bibfnamefont {L.~G.}\ \bibnamefont {Stanton}}, \bibinfo
  {author} {\bibfnamefont {C.~R.}\ \bibnamefont {Scullard}}, \bibinfo {author}
  {\bibfnamefont {A.~A.}\ \bibnamefont {Correa}}, \bibinfo {author}
  {\bibfnamefont {J.~I.}\ \bibnamefont {Castor}}, \bibinfo {author}
  {\bibfnamefont {F.~R.}\ \bibnamefont {Graziani}}, \bibinfo {author}
  {\bibfnamefont {L.~A.}\ \bibnamefont {Collins}}, \bibinfo {author}
  {\bibfnamefont {O.}~\bibnamefont {Certík}}, \bibinfo {author} {\bibfnamefont
  {J.~D.}\ \bibnamefont {Kress}}, \ and\ \bibinfo {author} {\bibfnamefont
  {M.~S.}\ \bibnamefont {Murillo}},\ }\bibfield  {title} {\enquote {\bibinfo
  {title} {Molecular dynamics studies of electron-ion temperature equilibration
  in hydrogen plasmas within the coupled-mode regime},}\ }\href
  {https://journals.aps.org/pre/abstract/10.1103/PhysRevE.95.043202} {\bibfield
   {journal} {\bibinfo  {journal} {Phys. Rev. E}\ }\textbf {\bibinfo {volume}
  {95}},\ \bibinfo {pages} {043202} (\bibinfo {year} {2017})}\BibitemShut
  {NoStop}%
\bibitem [{\citenamefont {Bohm}\ and\ \citenamefont {D.~Pines}(1952)}]{pines}%
  \BibitemOpen
  \bibfield  {author} {\bibinfo {author} {\bibfnamefont {D.}~\bibnamefont
  {Bohm}}\ and\ \bibinfo {author} {\bibfnamefont {A}~\bibnamefont {D.~Pines}},\
  }\bibfield  {title} {\enquote {\bibinfo {title} {Collective description of
  electron interactions: Ii. collective vs individual particle aspects of the
  interactions},}\ }\href
  {https://journals.aps.org/pr/abstract/10.1103/PhysRev.85.338} {\bibfield
  {journal} {\bibinfo  {journal} {Phys. Rev.}\ }\textbf {\bibinfo {volume}
  {85}},\ \bibinfo {pages} {338} (\bibinfo {year} {1952})}\BibitemShut
  {NoStop}%
\bibitem [{\citenamefont {Kugler}(1975)}]{kugler1}%
  \BibitemOpen
  \bibfield  {author} {\bibinfo {author} {\bibfnamefont {A.~A.}\ \bibnamefont
  {Kugler}},\ }\bibfield  {title} {\enquote {\bibinfo {title} {Theory of the
  local field correction in an electron gas},}\ }\href
  {http://link.springer.com/article/10.1007/BF01024183} {\bibfield  {journal}
  {\bibinfo  {journal} {J. Stat. Phys}\ }\textbf {\bibinfo {volume} {12}},\
  \bibinfo {pages} {35} (\bibinfo {year} {1975})}\BibitemShut {NoStop}%
\bibitem [{\citenamefont {Singwi}\ \emph {et~al.}(1968)\citenamefont {Singwi},
  \citenamefont {Tosi}, \citenamefont {Land},\ and\ \citenamefont
  {Sj\"olander}}]{stls_original}%
  \BibitemOpen
  \bibfield  {author} {\bibinfo {author} {\bibfnamefont {K.~S.}\ \bibnamefont
  {Singwi}}, \bibinfo {author} {\bibfnamefont {M.~P.}\ \bibnamefont {Tosi}},
  \bibinfo {author} {\bibfnamefont {R.~H.}\ \bibnamefont {Land}}, \ and\
  \bibinfo {author} {\bibfnamefont {A.}~\bibnamefont {Sj\"olander}},\
  }\bibfield  {title} {\enquote {\bibinfo {title} {Electron correlations at
  metallic densities},}\ }\href
  {http://link.aps.org/doi/10.1103/PhysRev.176.589} {\bibfield  {journal}
  {\bibinfo  {journal} {Phys. Rev}\ }\textbf {\bibinfo {volume} {176}},\
  \bibinfo {pages} {589} (\bibinfo {year} {1968})}\BibitemShut {NoStop}%
\bibitem [{\citenamefont {Vashishta}\ and\ \citenamefont
  {Singwi}(1972)}]{vs_original}%
  \BibitemOpen
  \bibfield  {author} {\bibinfo {author} {\bibfnamefont {P.}~\bibnamefont
  {Vashishta}}\ and\ \bibinfo {author} {\bibfnamefont {K.~S.}\ \bibnamefont
  {Singwi}},\ }\bibfield  {title} {\enquote {\bibinfo {title} {Electron
  correlations at metallic densities v},}\ }\href
  {http://link.aps.org/doi/10.1103/PhysRevB.6.875} {\bibfield  {journal}
  {\bibinfo  {journal} {Phys. Rev. B}\ }\textbf {\bibinfo {volume} {6}},\
  \bibinfo {pages} {875} (\bibinfo {year} {1972})}\BibitemShut {NoStop}%
\bibitem [{\citenamefont {Holas}\ and\ \citenamefont
  {Rahman}(1987)}]{dynamic_ii}%
  \BibitemOpen
  \bibfield  {author} {\bibinfo {author} {\bibfnamefont {A.}~\bibnamefont
  {Holas}}\ and\ \bibinfo {author} {\bibfnamefont {S.}~\bibnamefont {Rahman}},\
  }\bibfield  {title} {\enquote {\bibinfo {title} {Dynamic local-field factor
  of an electron liquid in the quantum versions of the
  {S}ingwi-{T}osi-{L}and-{S}j\"olander and {V}ashishta-{S}ingwi theories},}\
  }\href {https://journals.aps.org/prb/abstract/10.1103/PhysRevB.35.2720}
  {\bibfield  {journal} {\bibinfo  {journal} {Phys. Rev. B}\ }\textbf {\bibinfo
  {volume} {35}},\ \bibinfo {pages} {2720} (\bibinfo {year}
  {1987})}\BibitemShut {NoStop}%
\bibitem [{\citenamefont {Farid}\ \emph {et~al.}(1993)\citenamefont {Farid},
  \citenamefont {Heine}, \citenamefont {Engel},\ and\ \citenamefont
  {Robertson}}]{farid}%
  \BibitemOpen
  \bibfield  {author} {\bibinfo {author} {\bibfnamefont {B.}~\bibnamefont
  {Farid}}, \bibinfo {author} {\bibfnamefont {V.}~\bibnamefont {Heine}},
  \bibinfo {author} {\bibfnamefont {G.~E.}\ \bibnamefont {Engel}}, \ and\
  \bibinfo {author} {\bibfnamefont {I.~J.}\ \bibnamefont {Robertson}},\
  }\bibfield  {title} {\enquote {\bibinfo {title} {Extremal properties of the
  harris-foulkes functional and an improved screening calculation for the
  electron gas},}\ }\href {http://link.aps.org/doi/10.1103/PhysRevB.48.11602}
  {\bibfield  {journal} {\bibinfo  {journal} {Phys. Rev. B}\ }\textbf {\bibinfo
  {volume} {48}},\ \bibinfo {pages} {11602} (\bibinfo {year}
  {1993})}\BibitemShut {NoStop}%
\bibitem [{\citenamefont {Moroni}\ \emph {et~al.}(1992)\citenamefont {Moroni},
  \citenamefont {Ceperley},\ and\ \citenamefont {Senatore}}]{moroni}%
  \BibitemOpen
  \bibfield  {author} {\bibinfo {author} {\bibfnamefont {S.}~\bibnamefont
  {Moroni}}, \bibinfo {author} {\bibfnamefont {D.~M.}\ \bibnamefont
  {Ceperley}}, \ and\ \bibinfo {author} {\bibfnamefont {G.}~\bibnamefont
  {Senatore}},\ }\bibfield  {title} {\enquote {\bibinfo {title} {Static
  response from quantum {M}onte {C}arlo calculations},}\ }\href
  {https://journals.aps.org/prl/abstract/10.1103/PhysRevLett.69.1837}
  {\bibfield  {journal} {\bibinfo  {journal} {Phys. Rev. Lett}\ }\textbf
  {\bibinfo {volume} {69}},\ \bibinfo {pages} {1837} (\bibinfo {year}
  {1992})}\BibitemShut {NoStop}%
\bibitem [{\citenamefont {Moroni}\ \emph {et~al.}(1995)\citenamefont {Moroni},
  \citenamefont {Ceperley},\ and\ \citenamefont {Senatore}}]{moroni2}%
  \BibitemOpen
  \bibfield  {author} {\bibinfo {author} {\bibfnamefont {S.}~\bibnamefont
  {Moroni}}, \bibinfo {author} {\bibfnamefont {D.~M.}\ \bibnamefont
  {Ceperley}}, \ and\ \bibinfo {author} {\bibfnamefont {G.}~\bibnamefont
  {Senatore}},\ }\bibfield  {title} {\enquote {\bibinfo {title} {Static
  response and local field factor of the electron gas},}\ }\href
  {http://link.aps.org/doi/10.1103/PhysRevLett.75.689} {\bibfield  {journal}
  {\bibinfo  {journal} {Phys. Rev. Lett}\ }\textbf {\bibinfo {volume} {75}},\
  \bibinfo {pages} {689} (\bibinfo {year} {1995})}\BibitemShut {NoStop}%
\bibitem [{\citenamefont {Bowen}\ \emph {et~al.}(1994)\citenamefont {Bowen},
  \citenamefont {Sugiyama},\ and\ \citenamefont {Alder}}]{bowen2}%
  \BibitemOpen
  \bibfield  {author} {\bibinfo {author} {\bibfnamefont {C.}~\bibnamefont
  {Bowen}}, \bibinfo {author} {\bibfnamefont {G.}~\bibnamefont {Sugiyama}}, \
  and\ \bibinfo {author} {\bibfnamefont {B.~J.}\ \bibnamefont {Alder}},\
  }\bibfield  {title} {\enquote {\bibinfo {title} {Static dielectric response
  of the electron gas},}\ }\href
  {http://link.aps.org/doi/10.1103/PhysRevB.50.14838} {\bibfield  {journal}
  {\bibinfo  {journal} {Phys. Rev. B}\ }\textbf {\bibinfo {volume} {50}},\
  \bibinfo {pages} {14838} (\bibinfo {year} {1994})}\BibitemShut {NoStop}%
\bibitem [{\citenamefont {Corradini}\ \emph {et~al.}(1998)\citenamefont
  {Corradini}, \citenamefont {Sole}, \citenamefont {Onida},\ and\ \citenamefont
  {Palummo}}]{cdop}%
  \BibitemOpen
  \bibfield  {author} {\bibinfo {author} {\bibfnamefont {M.}~\bibnamefont
  {Corradini}}, \bibinfo {author} {\bibfnamefont {R.~Del}\ \bibnamefont
  {Sole}}, \bibinfo {author} {\bibfnamefont {G.}~\bibnamefont {Onida}}, \ and\
  \bibinfo {author} {\bibfnamefont {M.}~\bibnamefont {Palummo}},\ }\bibfield
  {title} {\enquote {\bibinfo {title} {Analytical expressions for the
  local-field factor $g(q)$ and the exchange-correlation kernel
  ${K}_{\mathrm{xc}}(r)$ of the homogeneous electron gas},}\ }\href
  {http://link.aps.org/doi/10.1103/PhysRevB.57.14569} {\bibfield  {journal}
  {\bibinfo  {journal} {Phys. Rev. B}\ }\textbf {\bibinfo {volume} {57}},\
  \bibinfo {pages} {14569} (\bibinfo {year} {1998})}\BibitemShut {NoStop}%
\bibitem [{\citenamefont {Tanaka}\ and\ \citenamefont {Ichimaru}(1986)}]{stls}%
  \BibitemOpen
  \bibfield  {author} {\bibinfo {author} {\bibfnamefont {S.}~\bibnamefont
  {Tanaka}}\ and\ \bibinfo {author} {\bibfnamefont {S.}~\bibnamefont
  {Ichimaru}},\ }\bibfield  {title} {\enquote {\bibinfo {title} {Thermodynamics
  and correlational properties of finite-temperature electron liquids in the
  {S}ingwi-{T}osi-{Land}-{S}j\"olander approximation},}\ }\href
  {http://journals.jps.jp/doi/abs/10.1143/JPSJ.55.2278} {\bibfield  {journal}
  {\bibinfo  {journal} {J. Phys. Soc. Jpn}\ }\textbf {\bibinfo {volume} {55}},\
  \bibinfo {pages} {2278--2289} (\bibinfo {year} {1986})}\BibitemShut {NoStop}%
\bibitem [{\citenamefont {Schweng}\ and\ \citenamefont
  {B\"ohm}(1993)}]{schweng}%
  \BibitemOpen
  \bibfield  {author} {\bibinfo {author} {\bibfnamefont {H.~K.}\ \bibnamefont
  {Schweng}}\ and\ \bibinfo {author} {\bibfnamefont {H.~M.}\ \bibnamefont
  {B\"ohm}},\ }\bibfield  {title} {\enquote {\bibinfo {title}
  {Finite-temperature electron correlations in the framework of a dynamic
  local-field correction},}\ }\href
  {https://journals.aps.org/prb/abstract/10.1103/PhysRevB.48.2037} {\bibfield
  {journal} {\bibinfo  {journal} {Phys. Rev. B}\ }\textbf {\bibinfo {volume}
  {48}},\ \bibinfo {pages} {2037} (\bibinfo {year} {1993})}\BibitemShut
  {NoStop}%
\bibitem [{\citenamefont {Perrot}\ and\ \citenamefont
  {Dharma-wardana}(2000)}]{perrot}%
  \BibitemOpen
  \bibfield  {author} {\bibinfo {author} {\bibfnamefont {F.}~\bibnamefont
  {Perrot}}\ and\ \bibinfo {author} {\bibfnamefont {M.~W.~C.}\ \bibnamefont
  {Dharma-wardana}},\ }\bibfield  {title} {\enquote {\bibinfo {title}
  {Spin-polarized electron liquid at arbitrary temperatures:
  Exchange-correlation energies, electron-distribution functions, and the
  static response functions},}\ }\href
  {https://journals.aps.org/prb/abstract/10.1103/PhysRevB.62.16536} {\bibfield
  {journal} {\bibinfo  {journal} {Phys. Rev. B}\ }\textbf {\bibinfo {volume}
  {62}},\ \bibinfo {pages} {16536} (\bibinfo {year} {2000})}\BibitemShut
  {NoStop}%
\bibitem [{\citenamefont {Stolzmann}\ and\ \citenamefont
  {R\"osler}(2001)}]{stolzmann}%
  \BibitemOpen
  \bibfield  {author} {\bibinfo {author} {\bibfnamefont {W.}~\bibnamefont
  {Stolzmann}}\ and\ \bibinfo {author} {\bibfnamefont {M.}~\bibnamefont
  {R\"osler}},\ }\bibfield  {title} {\enquote {\bibinfo {title} {Static
  local-field corrected dielectric and thermodynamic functions},}\ }\href
  {http://onlinelibrary.wiley.com/doi/10.1002/1521-3986(200103)41:2/3<203::AID-CTPP203>3.0.CO;2-S/abstract}
  {\bibfield  {journal} {\bibinfo  {journal} {Contrib. Plasma Phys}\ }\textbf
  {\bibinfo {volume} {41}},\ \bibinfo {pages} {203} (\bibinfo {year}
  {2001})}\BibitemShut {NoStop}%
\bibitem [{\citenamefont {Sjostrom}\ and\ \citenamefont {Dufty}(2013)}]{stls2}%
  \BibitemOpen
  \bibfield  {author} {\bibinfo {author} {\bibfnamefont {T.}~\bibnamefont
  {Sjostrom}}\ and\ \bibinfo {author} {\bibfnamefont {J.}~\bibnamefont
  {Dufty}},\ }\bibfield  {title} {\enquote {\bibinfo {title} {Uniform electron
  gas at finite temperatures},}\ }\href
  {http://link.aps.org/doi/10.1103/PhysRevB.88.115123} {\bibfield  {journal}
  {\bibinfo  {journal} {Phys. Rev. B}\ }\textbf {\bibinfo {volume} {88}},\
  \bibinfo {pages} {115123} (\bibinfo {year} {2013})}\BibitemShut {NoStop}%
\bibitem [{\citenamefont {Tanaka}(2016)}]{tanaka_hnc}%
  \BibitemOpen
  \bibfield  {author} {\bibinfo {author} {\bibfnamefont {S.}~\bibnamefont
  {Tanaka}},\ }\bibfield  {title} {\enquote {\bibinfo {title} {Correlational
  and thermodynamic properties of finite-temperature electron liquids in the
  hypernetted-chain approximation},}\ }\href
  {https://aip.scitation.org/doi/abs/10.1063/1.4969071} {\bibfield  {journal}
  {\bibinfo  {journal} {J. Chem. Phys}\ }\textbf {\bibinfo {volume} {145}},\
  \bibinfo {pages} {214104} (\bibinfo {year} {2016})}\BibitemShut {NoStop}%
\bibitem [{\citenamefont {Dornheim}\ \emph
  {et~al.}(2017{\natexlab{a}})\citenamefont {Dornheim}, \citenamefont {Groth},
  \citenamefont {Vorberger},\ and\ \citenamefont {Bonitz}}]{dornheim_pre}%
  \BibitemOpen
  \bibfield  {author} {\bibinfo {author} {\bibfnamefont {T.}~\bibnamefont
  {Dornheim}}, \bibinfo {author} {\bibfnamefont {S.}~\bibnamefont {Groth}},
  \bibinfo {author} {\bibfnamefont {J.}~\bibnamefont {Vorberger}}, \ and\
  \bibinfo {author} {\bibfnamefont {M.}~\bibnamefont {Bonitz}},\ }\bibfield
  {title} {\enquote {\bibinfo {title} {Permutation blocking path integral
  {M}onte {C}arlo approach to the static density response of the warm dense
  electron gas},}\ }\href
  {https://journals.aps.org/pre/abstract/10.1103/PhysRevE.96.023203} {\bibfield
   {journal} {\bibinfo  {journal} {Phys. Rev. E}\ }\textbf {\bibinfo {volume}
  {96}},\ \bibinfo {pages} {023203} (\bibinfo {year}
  {2017}{\natexlab{a}})}\BibitemShut {NoStop}%
\bibitem [{\citenamefont {Groth}\ \emph
  {et~al.}(2017{\natexlab{a}})\citenamefont {Groth}, \citenamefont {Dornheim},\
  and\ \citenamefont {Bonitz}}]{groth_jcp}%
  \BibitemOpen
  \bibfield  {author} {\bibinfo {author} {\bibfnamefont {S.}~\bibnamefont
  {Groth}}, \bibinfo {author} {\bibfnamefont {T.}~\bibnamefont {Dornheim}}, \
  and\ \bibinfo {author} {\bibfnamefont {M.}~\bibnamefont {Bonitz}},\
  }\bibfield  {title} {\enquote {\bibinfo {title} {Configuration path integral
  {M}onte {C}arlo approach to the static density response of the warm dense
  electron gas},}\ }\href {https://aip.scitation.org/doi/abs/10.1063/1.4999907}
  {\bibfield  {journal} {\bibinfo  {journal} {J. Chem. Phys}\ }\textbf
  {\bibinfo {volume} {147}},\ \bibinfo {pages} {164108} (\bibinfo {year}
  {2017}{\natexlab{a}})}\BibitemShut {NoStop}%
\bibitem [{\citenamefont {Arora}\ \emph {et~al.}(2017)\citenamefont {Arora},
  \citenamefont {Kumar},\ and\ \citenamefont {Moudgil}}]{arora}%
  \BibitemOpen
  \bibfield  {author} {\bibinfo {author} {\bibfnamefont {P.}~\bibnamefont
  {Arora}}, \bibinfo {author} {\bibfnamefont {K.}~\bibnamefont {Kumar}}, \ and\
  \bibinfo {author} {\bibfnamefont {R.~K.}\ \bibnamefont {Moudgil}},\
  }\bibfield  {title} {\enquote {\bibinfo {title} {Spin-resolved correlations
  in the warm-dense homogeneous electron gas},}\ }\href
  {https://link.springer.com/article/10.1140/epjb/e2017-70532-y} {\bibfield
  {journal} {\bibinfo  {journal} {Eur. Phys. J. B}\ }\textbf {\bibinfo {volume}
  {90}},\ \bibinfo {pages} {76} (\bibinfo {year} {2017})}\BibitemShut {NoStop}%
\bibitem [{\citenamefont {Dornheim}\ \emph {et~al.}(2019)\citenamefont
  {Dornheim}, \citenamefont {Vorberger}, \citenamefont {Groth}, \citenamefont
  {Hoffmann}, \citenamefont {Moldabekov},\ and\ \citenamefont
  {Bonitz}}]{dornheim_ML}%
  \BibitemOpen
  \bibfield  {author} {\bibinfo {author} {\bibfnamefont {T.}~\bibnamefont
  {Dornheim}}, \bibinfo {author} {\bibfnamefont {J.}~\bibnamefont {Vorberger}},
  \bibinfo {author} {\bibfnamefont {S.}~\bibnamefont {Groth}}, \bibinfo
  {author} {\bibfnamefont {N.}~\bibnamefont {Hoffmann}}, \bibinfo {author}
  {\bibfnamefont {Zh.A.}\ \bibnamefont {Moldabekov}}, \ and\ \bibinfo {author}
  {\bibfnamefont {M.}~\bibnamefont {Bonitz}},\ }\bibfield  {title} {\enquote
  {\bibinfo {title} {The static local field correction of the warm dense
  electron gas: An ab initio path integral {M}onte {C}arlo study and machine
  learning representation},}\ }\href
  {https://aip.scitation.org/doi/full/10.1063/1.5123013} {\bibfield  {journal}
  {\bibinfo  {journal} {J. Chem. Phys}\ }\textbf {\bibinfo {volume} {151}},\
  \bibinfo {pages} {194104} (\bibinfo {year} {2019})}\BibitemShut {NoStop}%
\bibitem [{\citenamefont {Dornheim}\ \emph {et~al.}(2020)\citenamefont
  {Dornheim}, \citenamefont {Sjostrom}, \citenamefont {Tanaka},\ and\
  \citenamefont {Vorberger}}]{dornheim_electron_liquid}%
  \BibitemOpen
  \bibfield  {author} {\bibinfo {author} {\bibfnamefont {T.}~\bibnamefont
  {Dornheim}}, \bibinfo {author} {\bibfnamefont {T.}~\bibnamefont {Sjostrom}},
  \bibinfo {author} {\bibfnamefont {S.}~\bibnamefont {Tanaka}}, \ and\ \bibinfo
  {author} {\bibfnamefont {J.}~\bibnamefont {Vorberger}},\ }\bibfield  {title}
  {\enquote {\bibinfo {title} {Strongly coupled electron liquid: ab initio path
  integral {M}onte {C}arlo simulations and dielectric theories},}\ }\href@noop
  {} {\bibfield  {journal} {\bibinfo  {journal} {Phys. Rev. B (in press)}\ }
  (\bibinfo {year} {2020})}\BibitemShut {NoStop}%
\bibitem [{\citenamefont {Dornheim}\ \emph {et~al.}()\citenamefont {Dornheim},
  \citenamefont {Moldabekov}, \citenamefont {Vorberger},\ and\ \citenamefont
  {Groth}}]{dornheim_HEDP}%
  \BibitemOpen
  \bibfield  {author} {\bibinfo {author} {\bibfnamefont {T.}~\bibnamefont
  {Dornheim}}, \bibinfo {author} {\bibfnamefont {Zh.A.}\ \bibnamefont
  {Moldabekov}}, \bibinfo {author} {\bibfnamefont {J.}~\bibnamefont
  {Vorberger}}, \ and\ \bibinfo {author} {\bibfnamefont {S.}~\bibnamefont
  {Groth}},\ }\bibfield  {title} {\enquote {\bibinfo {title} {Ab initio path
  integral {M}onte {C}arlo simulation of the uniform electron gas in the high
  energy density regime},}\ }\href {https://arxiv.org/abs/2003.00858} {\
  }\Eprint {http://arxiv.org/abs/2003:00858} {arXiv:2003:00858} \BibitemShut
  {NoStop}%
\bibitem [{\citenamefont {Dornheim}\ \emph
  {et~al.}(2018{\natexlab{a}})\citenamefont {Dornheim}, \citenamefont {Groth},\
  and\ \citenamefont {Bonitz}}]{review}%
  \BibitemOpen
  \bibfield  {author} {\bibinfo {author} {\bibfnamefont {T.}~\bibnamefont
  {Dornheim}}, \bibinfo {author} {\bibfnamefont {S.}~\bibnamefont {Groth}}, \
  and\ \bibinfo {author} {\bibfnamefont {M.}~\bibnamefont {Bonitz}},\
  }\bibfield  {title} {\enquote {\bibinfo {title} {The uniform electron gas at
  warm dense matter conditions},}\ }\href
  {https://www.sciencedirect.com/science/article/abs/pii/S0370157318300516}
  {\bibfield  {journal} {\bibinfo  {journal} {Phys. Reports}\ }\textbf
  {\bibinfo {volume} {744}},\ \bibinfo {pages} {1--86} (\bibinfo {year}
  {2018}{\natexlab{a}})}\BibitemShut {NoStop}%
\bibitem [{\citenamefont {Dornheim}\ \emph
  {et~al.}(2018{\natexlab{b}})\citenamefont {Dornheim}, \citenamefont {Groth},
  \citenamefont {Vorberger},\ and\ \citenamefont {Bonitz}}]{dornheim_dynamic}%
  \BibitemOpen
  \bibfield  {author} {\bibinfo {author} {\bibfnamefont {T.}~\bibnamefont
  {Dornheim}}, \bibinfo {author} {\bibfnamefont {S.}~\bibnamefont {Groth}},
  \bibinfo {author} {\bibfnamefont {J.}~\bibnamefont {Vorberger}}, \ and\
  \bibinfo {author} {\bibfnamefont {M.}~\bibnamefont {Bonitz}},\ }\bibfield
  {title} {\enquote {\bibinfo {title} {Ab initio path integral {M}onte {C}arlo
  results for the dynamic structure factor of correlated electrons: From the
  electron liquid to warm dense matter},}\ }\href
  {https://journals.aps.org/prl/abstract/10.1103/PhysRevLett.121.255001}
  {\bibfield  {journal} {\bibinfo  {journal} {Phys. Rev. Lett.}\ }\textbf
  {\bibinfo {volume} {121}},\ \bibinfo {pages} {255001} (\bibinfo {year}
  {2018}{\natexlab{b}})}\BibitemShut {NoStop}%
\bibitem [{\citenamefont {Groth}\ \emph {et~al.}(2019)\citenamefont {Groth},
  \citenamefont {Dornheim},\ and\ \citenamefont
  {Vorberger}}]{dynamic_folgepaper}%
  \BibitemOpen
  \bibfield  {author} {\bibinfo {author} {\bibfnamefont {S.}~\bibnamefont
  {Groth}}, \bibinfo {author} {\bibfnamefont {T.}~\bibnamefont {Dornheim}}, \
  and\ \bibinfo {author} {\bibfnamefont {J.}~\bibnamefont {Vorberger}},\
  }\bibfield  {title} {\enquote {\bibinfo {title} {Ab initio path integral
  {M}onte {C}arlo approach to the static and dynamic density response of the
  uniform electron gas},}\ }\href
  {https://link.aps.org/doi/10.1103/PhysRevB.99.235122} {\bibfield  {journal}
  {\bibinfo  {journal} {Phys. Rev. B}\ }\textbf {\bibinfo {volume} {99}},\
  \bibinfo {pages} {235122} (\bibinfo {year} {2019})}\BibitemShut {NoStop}%
\bibitem [{\citenamefont {Fletcher}\ \emph {et~al.}(2015)\citenamefont
  {Fletcher}, \citenamefont {Lee}, \citenamefont {D{\"o}ppner}, \citenamefont
  {Galtier}, \citenamefont {Nagler}, \citenamefont {Heimann}, \citenamefont
  {Fortmann}, \citenamefont {LePape}, \citenamefont {Ma}, \citenamefont
  {Millot}, \citenamefont {Pak}, \citenamefont {Turnbull}, \citenamefont
  {Chapman}, \citenamefont {Gericke}, \citenamefont {Vorberger}, \citenamefont
  {White}, \citenamefont {Gregori}, \citenamefont {Wei}, \citenamefont
  {Barbrel}, \citenamefont {Falcone}, \citenamefont {Kao}, \citenamefont
  {Nuhn}, \citenamefont {Welch}, \citenamefont {Zastrau}, \citenamefont
  {Neumayer}, \citenamefont {Hastings},\ and\ \citenamefont
  {Glenzer}}]{Fletcher2015}%
  \BibitemOpen
  \bibfield  {author} {\bibinfo {author} {\bibfnamefont {L.~B.}\ \bibnamefont
  {Fletcher}}, \bibinfo {author} {\bibfnamefont {H.~J.}\ \bibnamefont {Lee}},
  \bibinfo {author} {\bibfnamefont {T.}~\bibnamefont {D{\"o}ppner}}, \bibinfo
  {author} {\bibfnamefont {E.}~\bibnamefont {Galtier}}, \bibinfo {author}
  {\bibfnamefont {B.}~\bibnamefont {Nagler}}, \bibinfo {author} {\bibfnamefont
  {P.}~\bibnamefont {Heimann}}, \bibinfo {author} {\bibfnamefont
  {C.}~\bibnamefont {Fortmann}}, \bibinfo {author} {\bibfnamefont
  {S.}~\bibnamefont {LePape}}, \bibinfo {author} {\bibfnamefont
  {T.}~\bibnamefont {Ma}}, \bibinfo {author} {\bibfnamefont {M.}~\bibnamefont
  {Millot}}, \bibinfo {author} {\bibfnamefont {A.}~\bibnamefont {Pak}},
  \bibinfo {author} {\bibfnamefont {D.}~\bibnamefont {Turnbull}}, \bibinfo
  {author} {\bibfnamefont {D.~A.}\ \bibnamefont {Chapman}}, \bibinfo {author}
  {\bibfnamefont {D.~O.}\ \bibnamefont {Gericke}}, \bibinfo {author}
  {\bibfnamefont {J.}~\bibnamefont {Vorberger}}, \bibinfo {author}
  {\bibfnamefont {T.}~\bibnamefont {White}}, \bibinfo {author} {\bibfnamefont
  {G.}~\bibnamefont {Gregori}}, \bibinfo {author} {\bibfnamefont
  {M.}~\bibnamefont {Wei}}, \bibinfo {author} {\bibfnamefont {B.}~\bibnamefont
  {Barbrel}}, \bibinfo {author} {\bibfnamefont {R.~W.}\ \bibnamefont
  {Falcone}}, \bibinfo {author} {\bibfnamefont {C.-C.}\ \bibnamefont {Kao}},
  \bibinfo {author} {\bibfnamefont {H.}~\bibnamefont {Nuhn}}, \bibinfo {author}
  {\bibfnamefont {J.}~\bibnamefont {Welch}}, \bibinfo {author} {\bibfnamefont
  {U.}~\bibnamefont {Zastrau}}, \bibinfo {author} {\bibfnamefont
  {P.}~\bibnamefont {Neumayer}}, \bibinfo {author} {\bibfnamefont {J.~B.}\
  \bibnamefont {Hastings}}, \ and\ \bibinfo {author} {\bibfnamefont {S.~H.}\
  \bibnamefont {Glenzer}},\ }\bibfield  {title} {\enquote {\bibinfo {title}
  {Ultrabright x-ray laser scattering for dynamic warm dense matter physics},}\
  }\href {https://doi.org/10.1038/nphoton.2015.41} {\bibfield  {journal}
  {\bibinfo  {journal} {Nature Photonics}\ }\textbf {\bibinfo {volume} {9}},\
  \bibinfo {pages} {274--279} (\bibinfo {year} {2015})}\BibitemShut {NoStop}%
\bibitem [{\citenamefont {Zastrau}\ \emph {et~al.}(2014)\citenamefont
  {Zastrau}, \citenamefont {Sperling}, \citenamefont {Harmand}, \citenamefont
  {Becker}, \citenamefont {Bornath}, \citenamefont {Bredow}, \citenamefont
  {Dziarzhytski}, \citenamefont {Fennel}, \citenamefont {Fletcher},
  \citenamefont {F{"o}rster}, \citenamefont {G{"o}de}, \citenamefont {Gregori},
  \citenamefont {Hilbert}, \citenamefont {Hochhaus}, \citenamefont {Holst},
  \citenamefont {Laarmann}, \citenamefont {Lee}, \citenamefont {Ma},
  \citenamefont {Mithen}, \citenamefont {Mitzner}, \citenamefont {Murphy},
  \citenamefont {Nakatsutsumi}, \citenamefont {Neumayer}, \citenamefont
  {Przystawik}, \citenamefont {Roling}, \citenamefont {Schulz}, \citenamefont
  {Siemer}, \citenamefont {Skruszewicz}, \citenamefont {Tiggesb{"a}umker},
  \citenamefont {Toleikis}, \citenamefont {Tschentscher}, \citenamefont
  {White}, \citenamefont {W{"o}stmann}, \citenamefont {Zacharias},
  \citenamefont {D{"o}ppner}, \citenamefont {Glenzer},\ and\ \citenamefont
  {Redmer}}]{Zastrau}%
  \BibitemOpen
  \bibfield  {author} {\bibinfo {author} {\bibfnamefont {U.}~\bibnamefont
  {Zastrau}}, \bibinfo {author} {\bibfnamefont {P.}~\bibnamefont {Sperling}},
  \bibinfo {author} {\bibfnamefont {M.}~\bibnamefont {Harmand}}, \bibinfo
  {author} {\bibfnamefont {A.}~\bibnamefont {Becker}}, \bibinfo {author}
  {\bibfnamefont {T.}~\bibnamefont {Bornath}}, \bibinfo {author} {\bibfnamefont
  {R.}~\bibnamefont {Bredow}}, \bibinfo {author} {\bibfnamefont
  {S.}~\bibnamefont {Dziarzhytski}}, \bibinfo {author} {\bibfnamefont
  {T.}~\bibnamefont {Fennel}}, \bibinfo {author} {\bibfnamefont {L.~B.}\
  \bibnamefont {Fletcher}}, \bibinfo {author} {\bibfnamefont {E.}~\bibnamefont
  {F{"o}rster}}, \bibinfo {author} {\bibfnamefont {S.}~\bibnamefont {G{"o}de}},
  \bibinfo {author} {\bibfnamefont {G.}~\bibnamefont {Gregori}}, \bibinfo
  {author} {\bibfnamefont {V.}~\bibnamefont {Hilbert}}, \bibinfo {author}
  {\bibfnamefont {D.}~\bibnamefont {Hochhaus}}, \bibinfo {author}
  {\bibfnamefont {B.}~\bibnamefont {Holst}}, \bibinfo {author} {\bibfnamefont
  {T.}~\bibnamefont {Laarmann}}, \bibinfo {author} {\bibfnamefont {H.~J.}\
  \bibnamefont {Lee}}, \bibinfo {author} {\bibfnamefont {T.}~\bibnamefont
  {Ma}}, \bibinfo {author} {\bibfnamefont {J.~P.}\ \bibnamefont {Mithen}},
  \bibinfo {author} {\bibfnamefont {R.}~\bibnamefont {Mitzner}}, \bibinfo
  {author} {\bibfnamefont {C.~D.}\ \bibnamefont {Murphy}}, \bibinfo {author}
  {\bibfnamefont {M.}~\bibnamefont {Nakatsutsumi}}, \bibinfo {author}
  {\bibfnamefont {P.}~\bibnamefont {Neumayer}}, \bibinfo {author}
  {\bibfnamefont {A.}~\bibnamefont {Przystawik}}, \bibinfo {author}
  {\bibfnamefont {S.}~\bibnamefont {Roling}}, \bibinfo {author} {\bibfnamefont
  {M.}~\bibnamefont {Schulz}}, \bibinfo {author} {\bibfnamefont
  {B.}~\bibnamefont {Siemer}}, \bibinfo {author} {\bibfnamefont
  {S.}~\bibnamefont {Skruszewicz}}, \bibinfo {author} {\bibfnamefont
  {J.}~\bibnamefont {Tiggesb{"a}umker}}, \bibinfo {author} {\bibfnamefont
  {S.}~\bibnamefont {Toleikis}}, \bibinfo {author} {\bibfnamefont
  {T.}~\bibnamefont {Tschentscher}}, \bibinfo {author} {\bibfnamefont
  {T.}~\bibnamefont {White}}, \bibinfo {author} {\bibfnamefont
  {M.}~\bibnamefont {W{"o}stmann}}, \bibinfo {author} {\bibfnamefont
  {H.}~\bibnamefont {Zacharias}}, \bibinfo {author} {\bibfnamefont
  {T.}~\bibnamefont {D{"o}ppner}}, \bibinfo {author} {\bibfnamefont {S.~H.}\
  \bibnamefont {Glenzer}}, \ and\ \bibinfo {author} {\bibfnamefont
  {R.}~\bibnamefont {Redmer}},\ }\bibfield  {title} {\enquote {\bibinfo {title}
  {Resolving ultrafast heating of dense cryogenic hydrogen},}\ }\href
  {https://journals.aps.org/prl/abstract/10.1103/PhysRevLett.112.105002}
  {\bibfield  {journal} {\bibinfo  {journal} {Phys. Rev. Lett}\ }\textbf
  {\bibinfo {volume} {112}},\ \bibinfo {pages} {105002} (\bibinfo {year}
  {2014})}\BibitemShut {NoStop}%
\bibitem [{\citenamefont {Ofori-Okai}\ \emph {et~al.}(2018)\citenamefont
  {Ofori-Okai}, \citenamefont {Hoffmann}, \citenamefont {Reid}, \citenamefont
  {Edstrom}, \citenamefont {Jobe}, \citenamefont {Li}, \citenamefont
  {Mannebach}, \citenamefont {Park}, \citenamefont {Polzin}, \citenamefont
  {Shen}, \citenamefont {Weathersby}, \citenamefont {Yang}, \citenamefont
  {Zheng}, \citenamefont {Zajac}, \citenamefont {Lindenberg}, \citenamefont
  {Glenzer},\ and\ \citenamefont {Wang}}]{Ofori_Okai_2018}%
  \BibitemOpen
  \bibfield  {author} {\bibinfo {author} {\bibfnamefont {B.~K.}\ \bibnamefont
  {Ofori-Okai}}, \bibinfo {author} {\bibfnamefont {M.~C.}\ \bibnamefont
  {Hoffmann}}, \bibinfo {author} {\bibfnamefont {A.~H.}\ \bibnamefont {Reid}},
  \bibinfo {author} {\bibfnamefont {S.}~\bibnamefont {Edstrom}}, \bibinfo
  {author} {\bibfnamefont {R.~K.}\ \bibnamefont {Jobe}}, \bibinfo {author}
  {\bibfnamefont {R.~K.}\ \bibnamefont {Li}}, \bibinfo {author} {\bibfnamefont
  {E.~M.}\ \bibnamefont {Mannebach}}, \bibinfo {author} {\bibfnamefont {S.~J.}\
  \bibnamefont {Park}}, \bibinfo {author} {\bibfnamefont {W.}~\bibnamefont
  {Polzin}}, \bibinfo {author} {\bibfnamefont {X.}~\bibnamefont {Shen}},
  \bibinfo {author} {\bibfnamefont {S.~P.}\ \bibnamefont {Weathersby}},
  \bibinfo {author} {\bibfnamefont {J.}~\bibnamefont {Yang}}, \bibinfo {author}
  {\bibfnamefont {Q.}~\bibnamefont {Zheng}}, \bibinfo {author} {\bibfnamefont
  {M.}~\bibnamefont {Zajac}}, \bibinfo {author} {\bibfnamefont {A.~M.}\
  \bibnamefont {Lindenberg}}, \bibinfo {author} {\bibfnamefont {S.~H.}\
  \bibnamefont {Glenzer}}, \ and\ \bibinfo {author} {\bibfnamefont {X.~J.}\
  \bibnamefont {Wang}},\ }\bibfield  {title} {\enquote {\bibinfo {title} {A
  terahertz pump mega-electron-volt ultrafast electron diffraction probe
  apparatus at the {SLAC} accelerator structure test area facility},}\ }\href
  {https://doi.org/10.1088\%2F1748-0221\%2F13\%2F06\%2Fp06014} {\bibfield
  {journal} {\bibinfo  {journal} {J. Inst}\ }\textbf {\bibinfo {volume} {13}},\
  \bibinfo {pages} {P06014--P06014} (\bibinfo {year} {2018})}\BibitemShut
  {NoStop}%
\bibitem [{\citenamefont {Goulielmakis}\ \emph {et~al.}(2004)\citenamefont
  {Goulielmakis}, \citenamefont {Uiberacker}, \citenamefont {Kienberger},
  \citenamefont {Baltuska}, \citenamefont {Yakovlev}, \citenamefont {Scrinzi},
  \citenamefont {Westerwalbesloh}, \citenamefont {Kleineberg}, \citenamefont
  {Heinzmann}, \citenamefont {Drescher},\ and\ \citenamefont
  {Krausz}}]{Goulielmakis1267}%
  \BibitemOpen
  \bibfield  {author} {\bibinfo {author} {\bibfnamefont {E.}~\bibnamefont
  {Goulielmakis}}, \bibinfo {author} {\bibfnamefont {M.}~\bibnamefont
  {Uiberacker}}, \bibinfo {author} {\bibfnamefont {R.}~\bibnamefont
  {Kienberger}}, \bibinfo {author} {\bibfnamefont {A.}~\bibnamefont
  {Baltuska}}, \bibinfo {author} {\bibfnamefont {V.}~\bibnamefont {Yakovlev}},
  \bibinfo {author} {\bibfnamefont {A.}~\bibnamefont {Scrinzi}}, \bibinfo
  {author} {\bibfnamefont {T.}~\bibnamefont {Westerwalbesloh}}, \bibinfo
  {author} {\bibfnamefont {U.}~\bibnamefont {Kleineberg}}, \bibinfo {author}
  {\bibfnamefont {U.}~\bibnamefont {Heinzmann}}, \bibinfo {author}
  {\bibfnamefont {M.}~\bibnamefont {Drescher}}, \ and\ \bibinfo {author}
  {\bibfnamefont {F.}~\bibnamefont {Krausz}},\ }\bibfield  {title} {\enquote
  {\bibinfo {title} {Direct measurement of light waves},}\ }\href
  {https://science.sciencemag.org/content/305/5688/1267.full.pdf} {\bibfield
  {journal} {\bibinfo  {journal} {Science}\ }\textbf {\bibinfo {volume}
  {305}},\ \bibinfo {pages} {1267--1269} (\bibinfo {year} {2004})}\BibitemShut
  {NoStop}%
\bibitem [{\citenamefont {Fr{\"u}hling}\ \emph {et~al.}(2009)\citenamefont
  {Fr{\"u}hling}, \citenamefont {Wieland}, \citenamefont {Gensch},
  \citenamefont {Gebert}, \citenamefont {Sch{\"u}tte}, \citenamefont
  {Krikunova}, \citenamefont {Kalms}, \citenamefont {Budzyn}, \citenamefont
  {Grimm}, \citenamefont {Rossbach}, \citenamefont {Pl{\"o}njes},\ and\
  \citenamefont {Drescher}}]{fruehling_np_09}%
  \BibitemOpen
  \bibfield  {author} {\bibinfo {author} {\bibfnamefont {U.}~\bibnamefont
  {Fr{\"u}hling}}, \bibinfo {author} {\bibfnamefont {M.}~\bibnamefont
  {Wieland}}, \bibinfo {author} {\bibfnamefont {M.}~\bibnamefont {Gensch}},
  \bibinfo {author} {\bibfnamefont {T.}~\bibnamefont {Gebert}}, \bibinfo
  {author} {\bibfnamefont {B.}~\bibnamefont {Sch{\"u}tte}}, \bibinfo {author}
  {\bibfnamefont {M.}~\bibnamefont {Krikunova}}, \bibinfo {author}
  {\bibfnamefont {R.}~\bibnamefont {Kalms}}, \bibinfo {author} {\bibfnamefont
  {F.}~\bibnamefont {Budzyn}}, \bibinfo {author} {\bibfnamefont
  {O.}~\bibnamefont {Grimm}}, \bibinfo {author} {\bibfnamefont
  {J.}~\bibnamefont {Rossbach}}, \bibinfo {author} {\bibfnamefont
  {E.}~\bibnamefont {Pl{\"o}njes}}, \ and\ \bibinfo {author} {\bibfnamefont
  {M.}~\bibnamefont {Drescher}},\ }\bibfield  {title} {\enquote {\bibinfo
  {title} {Single-shot terahertz-field-driven x-ray streak camera},}\ }\href
  {https://www.nature.com/articles/nphoton.2009.160} {\bibfield  {journal}
  {\bibinfo  {journal} {Nature Photonics}\ }\textbf {\bibinfo {volume} {3}},\
  \bibinfo {pages} {523} (\bibinfo {year} {2009})}\BibitemShut {NoStop}%
\bibitem [{\citenamefont {Kazansky}\ \emph {et~al.}(2019)\citenamefont
  {Kazansky}, \citenamefont {Sazhina},\ and\ \citenamefont
  {Kabachnik}}]{Kazansky_2019}%
  \BibitemOpen
  \bibfield  {author} {\bibinfo {author} {\bibfnamefont {A.~K.}\ \bibnamefont
  {Kazansky}}, \bibinfo {author} {\bibfnamefont {I.~P.}\ \bibnamefont
  {Sazhina}}, \ and\ \bibinfo {author} {\bibfnamefont {N.~M.}\ \bibnamefont
  {Kabachnik}},\ }\bibfield  {title} {\enquote {\bibinfo {title} {Angular
  streaking of auger-electrons by {THz} field},}\ }\href
  {https://iopscience.iop.org/article/10.1088/1361-6455/aafa33/meta} {\bibfield
   {journal} {\bibinfo  {journal} {J. Phys. B: Atomic, Mol. and Opt. Phys}\
  }\textbf {\bibinfo {volume} {52}},\ \bibinfo {pages} {045601} (\bibinfo
  {year} {2019})}\BibitemShut {NoStop}%
\bibitem [{\citenamefont {Golden}\ \emph {et~al.}(1985)\citenamefont {Golden},
  \citenamefont {Green},\ and\ \citenamefont {Neilson}}]{golden_pra_85}%
  \BibitemOpen
  \bibfield  {author} {\bibinfo {author} {\bibfnamefont {K.~I.}\ \bibnamefont
  {Golden}}, \bibinfo {author} {\bibfnamefont {F.}~\bibnamefont {Green}}, \
  and\ \bibinfo {author} {\bibfnamefont {D.}~\bibnamefont {Neilson}},\
  }\bibfield  {title} {\enquote {\bibinfo {title} {Nonlinear-response-function
  approach to binary ionic mixtures: Dynamical theory},}\ }\href
  {https://journals.aps.org/pra/abstract/10.1103/PhysRevA.32.1669} {\bibfield
  {journal} {\bibinfo  {journal} {Phys. Rev. A}\ }\textbf {\bibinfo {volume}
  {32}},\ \bibinfo {pages} {1669} (\bibinfo {year} {1985})}\BibitemShut
  {NoStop}%
\bibitem [{\citenamefont {Bonitz}\ \emph {et~al.}(2010)\citenamefont {Bonitz},
  \citenamefont {Donk\'o}, \citenamefont {Ott}, \citenamefont {K\"ahlert},\
  and\ \citenamefont {Hartmann}}]{bonitz_prl_12}%
  \BibitemOpen
  \bibfield  {author} {\bibinfo {author} {\bibfnamefont {M.}~\bibnamefont
  {Bonitz}}, \bibinfo {author} {\bibfnamefont {Z.}~\bibnamefont {Donk\'o}},
  \bibinfo {author} {\bibfnamefont {T.}~\bibnamefont {Ott}}, \bibinfo {author}
  {\bibfnamefont {H.}~\bibnamefont {K\"ahlert}}, \ and\ \bibinfo {author}
  {\bibfnamefont {P.}~\bibnamefont {Hartmann}},\ }\bibfield  {title} {\enquote
  {\bibinfo {title} {Nonlinear magnetoplasmons in strongly coupled yukawa
  plasmas},}\ }\href {https://link.aps.org/doi/10.1103/PhysRevLett.105.055002}
  {\bibfield  {journal} {\bibinfo  {journal} {Phys. Rev. Lett}\ }\textbf
  {\bibinfo {volume} {105}},\ \bibinfo {pages} {055002} (\bibinfo {year}
  {2010})}\BibitemShut {NoStop}%
\bibitem [{\citenamefont {Kwong}\ and\ \citenamefont
  {Bonitz}(2000)}]{kwong_prl-00}%
  \BibitemOpen
  \bibfield  {author} {\bibinfo {author} {\bibfnamefont {N.-H.}\ \bibnamefont
  {Kwong}}\ and\ \bibinfo {author} {\bibfnamefont {M.}~\bibnamefont {Bonitz}},\
  }\bibfield  {title} {\enquote {\bibinfo {title} {Real-time kadanoff-baym
  approach to plasma oscillations in a correlated electron gas},}\ }\href
  {https://journals.aps.org/prl/abstract/10.1103/PhysRevLett.84.1768}
  {\bibfield  {journal} {\bibinfo  {journal} {Phys. Rev. Lett}\ }\textbf
  {\bibinfo {volume} {84}},\ \bibinfo {pages} {1768} (\bibinfo {year}
  {2000})}\BibitemShut {NoStop}%
\bibitem [{\citenamefont {Haberland}\ \emph {et~al.}(2001)\citenamefont
  {Haberland}, \citenamefont {Bonitz},\ and\ \citenamefont
  {Kremp}}]{haberland_pre_01}%
  \BibitemOpen
  \bibfield  {author} {\bibinfo {author} {\bibfnamefont {H.}~\bibnamefont
  {Haberland}}, \bibinfo {author} {\bibfnamefont {M.}~\bibnamefont {Bonitz}}, \
  and\ \bibinfo {author} {\bibfnamefont {D.}~\bibnamefont {Kremp}},\ }\bibfield
   {title} {\enquote {\bibinfo {title} {Harmonics generation in electron-ion
  collisions in a short laser pulse},}\ }\href
  {https://link.aps.org/doi/10.1103/PhysRevE.64.026405} {\bibfield  {journal}
  {\bibinfo  {journal} {Phys. Rev. E}\ }\textbf {\bibinfo {volume} {64}},\
  \bibinfo {pages} {026405} (\bibinfo {year} {2001})}\BibitemShut {NoStop}%
\bibitem [{\citenamefont {Nolting}\ and\ \citenamefont
  {Brewer}(2009)}]{nolting}%
  \BibitemOpen
  \bibfield  {author} {\bibinfo {author} {\bibfnamefont {W.}~\bibnamefont
  {Nolting}}\ and\ \bibinfo {author} {\bibfnamefont {W.~D.}\ \bibnamefont
  {Brewer}},\ }\href@noop {} {\emph {\bibinfo {title} {Fundamentals of
  Many-body Physics: Principles and Methods}}}\ (\bibinfo  {publisher}
  {Springer},\ \bibinfo {address} {Heidelberg},\ \bibinfo {year}
  {2009})\BibitemShut {NoStop}%
\bibitem [{\citenamefont {Giuliani}\ and\ \citenamefont
  {Vignale}(2008)}]{quantum_theory}%
  \BibitemOpen
  \bibfield  {author} {\bibinfo {author} {\bibfnamefont {G.}~\bibnamefont
  {Giuliani}}\ and\ \bibinfo {author} {\bibfnamefont {G.}~\bibnamefont
  {Vignale}},\ }\href@noop {} {\emph {\bibinfo {title} {Quantum Theory of the
  Electron Liquid}}}\ (\bibinfo  {publisher} {Cambridge University Press},\
  \bibinfo {address} {Cambridge},\ \bibinfo {year} {2008})\BibitemShut
  {NoStop}%
\bibitem [{\citenamefont {Baroni}\ \emph {et~al.}(1987)\citenamefont {Baroni},
  \citenamefont {Gianozzi},\ and\ \citenamefont {Testa}}]{baroni1}%
  \BibitemOpen
  \bibfield  {author} {\bibinfo {author} {\bibfnamefont {S.}~\bibnamefont
  {Baroni}}, \bibinfo {author} {\bibfnamefont {P.}~\bibnamefont {Gianozzi}}, \
  and\ \bibinfo {author} {\bibfnamefont {A.}~\bibnamefont {Testa}},\ }\bibfield
   {title} {\enquote {\bibinfo {title} {Green's-function approach to linear
  response in solids},}\ }\href
  {https://journals.aps.org/prl/abstract/10.1103/PhysRevLett.58.1861}
  {\bibfield  {journal} {\bibinfo  {journal} {Phys. Rev. Lett}\ }\textbf
  {\bibinfo {volume} {58}},\ \bibinfo {pages} {1861} (\bibinfo {year}
  {1987})}\BibitemShut {NoStop}%
\bibitem [{\citenamefont {Baroni}\ \emph {et~al.}(2001)\citenamefont {Baroni},
  \citenamefont {de~Gironcoli}, \citenamefont {Corso},\ and\ \citenamefont
  {Gianozzi}}]{baroni2}%
  \BibitemOpen
  \bibfield  {author} {\bibinfo {author} {\bibfnamefont {S.}~\bibnamefont
  {Baroni}}, \bibinfo {author} {\bibfnamefont {S.}~\bibnamefont
  {de~Gironcoli}}, \bibinfo {author} {\bibfnamefont {A.~Dal}\ \bibnamefont
  {Corso}}, \ and\ \bibinfo {author} {\bibfnamefont {P.}~\bibnamefont
  {Gianozzi}},\ }\bibfield  {title} {\enquote {\bibinfo {title} {Phonons and
  related crystal properties from density-functionalperturbation theory},}\
  }\href {https://journals.aps.org/rmp/abstract/10.1103/RevModPhys.73.515}
  {\bibfield  {journal} {\bibinfo  {journal} {Rev. Mod. Phys}\ }\textbf
  {\bibinfo {volume} {73}},\ \bibinfo {pages} {515} (\bibinfo {year}
  {2001})}\BibitemShut {NoStop}%
\bibitem [{\citenamefont {Haussmann}\ \emph {et~al.}(2009)\citenamefont
  {Haussmann}, \citenamefont {Punk},\ and\ \citenamefont
  {Zwerger}}]{ultracold1}%
  \BibitemOpen
  \bibfield  {author} {\bibinfo {author} {\bibfnamefont {R.}~\bibnamefont
  {Haussmann}}, \bibinfo {author} {\bibfnamefont {M.}~\bibnamefont {Punk}}, \
  and\ \bibinfo {author} {\bibfnamefont {W.}~\bibnamefont {Zwerger}},\
  }\bibfield  {title} {\enquote {\bibinfo {title} {Spectral functions and rf
  response of ultracold fermionic atoms},}\ }\href
  {https://journals.aps.org/pra/abstract/10.1103/PhysRevA.80.063612} {\bibfield
   {journal} {\bibinfo  {journal} {Phys. Rev. A}\ }\textbf {\bibinfo {volume}
  {80}},\ \bibinfo {pages} {063612} (\bibinfo {year} {2009})}\BibitemShut
  {NoStop}%
\bibitem [{\citenamefont {Pollock}\ and\ \citenamefont
  {Ceperley}(1987)}]{ultracold2}%
  \BibitemOpen
  \bibfield  {author} {\bibinfo {author} {\bibfnamefont {E.~L.}\ \bibnamefont
  {Pollock}}\ and\ \bibinfo {author} {\bibfnamefont {D.~M.}\ \bibnamefont
  {Ceperley}},\ }\bibfield  {title} {\enquote {\bibinfo {title} {Path-integral
  computation of superfluid densities},}\ }\href
  {https://journals.aps.org/prb/abstract/10.1103/PhysRevB.36.8343} {\bibfield
  {journal} {\bibinfo  {journal} {Phys. Rev. Lett}\ }\textbf {\bibinfo {volume}
  {36}},\ \bibinfo {pages} {8343} (\bibinfo {year} {1987})}\BibitemShut
  {NoStop}%
\bibitem [{\citenamefont {Nishikawa}\ and\ \citenamefont
  {Wakatani}(2000)}]{plasma1}%
  \BibitemOpen
  \bibfield  {author} {\bibinfo {author} {\bibfnamefont {K.}~\bibnamefont
  {Nishikawa}}\ and\ \bibinfo {author} {\bibfnamefont {M.}~\bibnamefont
  {Wakatani}},\ }\href@noop {} {\emph {\bibinfo {title} {Plasma Physics: Basic
  Theory with Fusion Applications}}}\ (\bibinfo  {publisher} {Springer Science
  \& Business Media},\ \bibinfo {address} {Heidelberg},\ \bibinfo {year}
  {2000})\BibitemShut {NoStop}%
\bibitem [{\citenamefont {Ichimaru}(1982)}]{plasma2}%
  \BibitemOpen
  \bibfield  {author} {\bibinfo {author} {\bibfnamefont {S.}~\bibnamefont
  {Ichimaru}},\ }\bibfield  {title} {\enquote {\bibinfo {title} {Strongly
  coupled plasmas: high-density classical plasmas and degenerate electron
  liquids},}\ }\href
  {https://journals.aps.org/rmp/abstract/10.1103/RevModPhys.54.1017} {\bibfield
   {journal} {\bibinfo  {journal} {Rev. Mod. Phys}\ }\textbf {\bibinfo {volume}
  {54}},\ \bibinfo {pages} {1017} (\bibinfo {year} {1982})}\BibitemShut
  {NoStop}%
\bibitem [{\citenamefont {Ceperley}(1995)}]{cep}%
  \BibitemOpen
  \bibfield  {author} {\bibinfo {author} {\bibfnamefont {D.~M.}\ \bibnamefont
  {Ceperley}},\ }\bibfield  {title} {\enquote {\bibinfo {title} {Path integrals
  in the theory of condensed helium},}\ }\href
  {https://journals.aps.org/rmp/abstract/10.1103/RevModPhys.67.279} {\bibfield
  {journal} {\bibinfo  {journal} {Rev. Mod. Phys}\ }\textbf {\bibinfo {volume}
  {67}},\ \bibinfo {pages} {279} (\bibinfo {year} {1995})}\BibitemShut
  {NoStop}%
\bibitem [{\citenamefont {Bonitz}(2016)}]{bonitz_book}%
  \BibitemOpen
  \bibfield  {author} {\bibinfo {author} {\bibfnamefont {M.}~\bibnamefont
  {Bonitz}},\ }\href@noop {} {\emph {\bibinfo {title} {Quantum kinetic
  theory}}}\ (\bibinfo  {publisher} {Springer},\ \bibinfo {address}
  {Heidelberg},\ \bibinfo {year} {2016})\BibitemShut {NoStop}%
\bibitem [{\citenamefont {Gross}\ and\ \citenamefont {Kohn}(1985)}]{dynamic1}%
  \BibitemOpen
  \bibfield  {author} {\bibinfo {author} {\bibfnamefont {E.~K.~U.}\
  \bibnamefont {Gross}}\ and\ \bibinfo {author} {\bibfnamefont
  {W.}~\bibnamefont {Kohn}},\ }\bibfield  {title} {\enquote {\bibinfo {title}
  {Local density-functional theory of frequency-dependent linear response},}\
  }\href {https://journals.aps.org/prl/abstract/10.1103/PhysRevLett.55.2850}
  {\bibfield  {journal} {\bibinfo  {journal} {Phys. Rev. Lett}\ }\textbf
  {\bibinfo {volume} {55}},\ \bibinfo {pages} {2850} (\bibinfo {year}
  {1985})}\BibitemShut {NoStop}%
\bibitem [{sup()}]{supplement}%
  \BibitemOpen
  \href@noop {} {}\bibinfo {note} {See Supplemental Material}\BibitemShut
  {NoStop}%
\bibitem [{\citenamefont {Loos}\ and\ \citenamefont {Gill}(2016)}]{loos}%
  \BibitemOpen
  \bibfield  {author} {\bibinfo {author} {\bibfnamefont {P.-F.}\ \bibnamefont
  {Loos}}\ and\ \bibinfo {author} {\bibfnamefont {P.~M.~W.}\ \bibnamefont
  {Gill}},\ }\bibfield  {title} {\enquote {\bibinfo {title} {The uniform
  electron gas},}\ }\href
  {http://onlinelibrary.wiley.com/doi/10.1002/wcms.1257/abstract} {\bibfield
  {journal} {\bibinfo  {journal} {Comput. Mol. Sci}\ }\textbf {\bibinfo
  {volume} {6}},\ \bibinfo {pages} {410--429} (\bibinfo {year}
  {2016})}\BibitemShut {NoStop}%
\bibitem [{\citenamefont {Mezzacapo}\ and\ \citenamefont
  {Boninsegni}(2007)}]{mezza}%
  \BibitemOpen
  \bibfield  {author} {\bibinfo {author} {\bibfnamefont {F.}~\bibnamefont
  {Mezzacapo}}\ and\ \bibinfo {author} {\bibfnamefont {M.}~\bibnamefont
  {Boninsegni}},\ }\bibfield  {title} {\enquote {\bibinfo {title} {Structure,
  superfluidity, and quantum melting of hydrogen clusters},}\ }\href
  {https://journals.aps.org/pra/abstract/10.1103/PhysRevA.75.033201} {\bibfield
   {journal} {\bibinfo  {journal} {Phys. Rev. A}\ }\textbf {\bibinfo {volume}
  {75}},\ \bibinfo {pages} {033201} (\bibinfo {year} {2007})}\BibitemShut
  {NoStop}%
\bibitem [{\citenamefont {Boninsegni}\ \emph
  {et~al.}(2006{\natexlab{a}})\citenamefont {Boninsegni}, \citenamefont
  {Prokofev},\ and\ \citenamefont {Svistunov}}]{boninsegni1}%
  \BibitemOpen
  \bibfield  {author} {\bibinfo {author} {\bibfnamefont {M.}~\bibnamefont
  {Boninsegni}}, \bibinfo {author} {\bibfnamefont {N.~V.}\ \bibnamefont
  {Prokofev}}, \ and\ \bibinfo {author} {\bibfnamefont {B.~V.}\ \bibnamefont
  {Svistunov}},\ }\bibfield  {title} {\enquote {\bibinfo {title} {Worm
  algorithm and diagrammatic {M}onte {C}arlo: A new approach to
  continuous-space path integral {M}onte {C}arlo simulations},}\ }\href
  {https://journals.aps.org/pre/abstract/10.1103/PhysRevE.74.036701} {\bibfield
   {journal} {\bibinfo  {journal} {Phys. Rev. E}\ }\textbf {\bibinfo {volume}
  {74}},\ \bibinfo {pages} {036701} (\bibinfo {year}
  {2006}{\natexlab{a}})}\BibitemShut {NoStop}%
\bibitem [{\citenamefont {Boninsegni}\ \emph
  {et~al.}(2006{\natexlab{b}})\citenamefont {Boninsegni}, \citenamefont
  {Prokofev},\ and\ \citenamefont {Svistunov}}]{boninsegni2}%
  \BibitemOpen
  \bibfield  {author} {\bibinfo {author} {\bibfnamefont {M.}~\bibnamefont
  {Boninsegni}}, \bibinfo {author} {\bibfnamefont {N.~V.}\ \bibnamefont
  {Prokofev}}, \ and\ \bibinfo {author} {\bibfnamefont {B.~V.}\ \bibnamefont
  {Svistunov}},\ }\bibfield  {title} {\enquote {\bibinfo {title} {Worm
  algorithm for continuous-space path integral {M}onte {C}arlo simulations},}\
  }\href {https://journals.aps.org/prl/abstract/10.1103/PhysRevLett.96.070601}
  {\bibfield  {journal} {\bibinfo  {journal} {Phys. Rev. Lett}\ }\textbf
  {\bibinfo {volume} {96}},\ \bibinfo {pages} {070601} (\bibinfo {year}
  {2006}{\natexlab{b}})}\BibitemShut {NoStop}%
\bibitem [{\citenamefont {Troyer}\ and\ \citenamefont {Wiese}(2005)}]{troyer}%
  \BibitemOpen
  \bibfield  {author} {\bibinfo {author} {\bibfnamefont {M.}~\bibnamefont
  {Troyer}}\ and\ \bibinfo {author} {\bibfnamefont {U.~J.}\ \bibnamefont
  {Wiese}},\ }\bibfield  {title} {\enquote {\bibinfo {title} {Computational
  complexity and fundamental limitations to fermionic quantum {M}onte {C}arlo
  simulations},}\ }\href
  {http://link.aps.org/doi/10.1103/PhysRevLett.94.170201} {\bibfield  {journal}
  {\bibinfo  {journal} {Phys. Rev. Lett}\ }\textbf {\bibinfo {volume} {94}},\
  \bibinfo {pages} {170201} (\bibinfo {year} {2005})}\BibitemShut {NoStop}%
\bibitem [{\citenamefont {Dornheim}(2019)}]{dornheim_sign_problem}%
  \BibitemOpen
  \bibfield  {author} {\bibinfo {author} {\bibfnamefont {T.}~\bibnamefont
  {Dornheim}},\ }\bibfield  {title} {\enquote {\bibinfo {title} {Fermion sign
  problem in path integral {M}onte {C}arlo simulations: Quantum dots, ultracold
  atoms, and warm dense matter},}\ }\href
  {https://journals.aps.org/pre/abstract/10.1103/PhysRevE.100.023307}
  {\bibfield  {journal} {\bibinfo  {journal} {Phys. Rev. E}\ }\textbf {\bibinfo
  {volume} {100}},\ \bibinfo {pages} {023307} (\bibinfo {year}
  {2019})}\BibitemShut {NoStop}%
\bibitem [{\citenamefont {Dornheim}\ \emph {et~al.}(2016)\citenamefont
  {Dornheim}, \citenamefont {Groth}, \citenamefont {Sjostrom}, \citenamefont
  {Malone}, \citenamefont {Foulkes},\ and\ \citenamefont
  {Bonitz}}]{dornheim_prl}%
  \BibitemOpen
  \bibfield  {author} {\bibinfo {author} {\bibfnamefont {T.}~\bibnamefont
  {Dornheim}}, \bibinfo {author} {\bibfnamefont {S.}~\bibnamefont {Groth}},
  \bibinfo {author} {\bibfnamefont {T.}~\bibnamefont {Sjostrom}}, \bibinfo
  {author} {\bibfnamefont {F.~D.}\ \bibnamefont {Malone}}, \bibinfo {author}
  {\bibfnamefont {W.~M.~C.}\ \bibnamefont {Foulkes}}, \ and\ \bibinfo {author}
  {\bibfnamefont {M.}~\bibnamefont {Bonitz}},\ }\bibfield  {title} {\enquote
  {\bibinfo {title} {Ab initio quantum {M}onte {C}arlo simulation of the warm
  dense electron gas in the thermodynamic limit},}\ }\href
  {http://link.aps.org/doi/10.1103/PhysRevLett.117.156403} {\bibfield
  {journal} {\bibinfo  {journal} {Phys. Rev. Lett.}\ }\textbf {\bibinfo
  {volume} {117}},\ \bibinfo {pages} {156403} (\bibinfo {year}
  {2016})}\BibitemShut {NoStop}%
\bibitem [{\citenamefont {Dornheim}\ \emph
  {et~al.}(2017{\natexlab{b}})\citenamefont {Dornheim}, \citenamefont {Groth},\
  and\ \citenamefont {Bonitz}}]{dornheim_cpp}%
  \BibitemOpen
  \bibfield  {author} {\bibinfo {author} {\bibfnamefont {T.}~\bibnamefont
  {Dornheim}}, \bibinfo {author} {\bibfnamefont {S.}~\bibnamefont {Groth}}, \
  and\ \bibinfo {author} {\bibfnamefont {M.}~\bibnamefont {Bonitz}},\
  }\bibfield  {title} {\enquote {\bibinfo {title} {Ab initio results for the
  static structure factor of the warm dense electron gas},}\ }\href
  {https://onlinelibrary.wiley.com/doi/full/10.1002/ctpp.201700096} {\bibfield
  {journal} {\bibinfo  {journal} {Contrib. Plasma Phys}\ }\textbf {\bibinfo
  {volume} {57}},\ \bibinfo {pages} {468--478} (\bibinfo {year}
  {2017}{\natexlab{b}})}\BibitemShut {NoStop}%
\bibitem [{\citenamefont {Kugler}(1970)}]{kugler_bounds}%
  \BibitemOpen
  \bibfield  {author} {\bibinfo {author} {\bibfnamefont {A.~A.}\ \bibnamefont
  {Kugler}},\ }\bibfield  {title} {\enquote {\bibinfo {title} {Bounds for some
  equilibrium properties of an electron gas},}\ }\href
  {https://journals.aps.org/pra/abstract/10.1103/PhysRevA.1.1688} {\bibfield
  {journal} {\bibinfo  {journal} {Phys. Rev. A}\ }\textbf {\bibinfo {volume}
  {1}},\ \bibinfo {pages} {1688} (\bibinfo {year} {1970})}\BibitemShut
  {NoStop}%
\bibitem [{\citenamefont {Benage}\ \emph {et~al.}(1999)\citenamefont {Benage},
  \citenamefont {Shanahan},\ and\ \citenamefont {Murillo}}]{benage}%
  \BibitemOpen
  \bibfield  {author} {\bibinfo {author} {\bibfnamefont {J.~F.}\ \bibnamefont
  {Benage}}, \bibinfo {author} {\bibfnamefont {W.~R.}\ \bibnamefont
  {Shanahan}}, \ and\ \bibinfo {author} {\bibfnamefont {M.~S.}\ \bibnamefont
  {Murillo}},\ }\bibfield  {title} {\enquote {\bibinfo {title} {Electrical
  resistivity measurements of hot dense aluminum},}\ }\href
  {https://journals.aps.org/prl/abstract/10.1103/PhysRevLett.83.2953}
  {\bibfield  {journal} {\bibinfo  {journal} {Phys. Rev. Lett}\ }\textbf
  {\bibinfo {volume} {83}},\ \bibinfo {pages} {2953} (\bibinfo {year}
  {1999})}\BibitemShut {NoStop}%
\bibitem [{\citenamefont {Karasiev}\ \emph {et~al.}(2016)\citenamefont
  {Karasiev}, \citenamefont {Calderin},\ and\ \citenamefont
  {Trickey}}]{karasiev_importance}%
  \BibitemOpen
  \bibfield  {author} {\bibinfo {author} {\bibfnamefont {V.~V.}\ \bibnamefont
  {Karasiev}}, \bibinfo {author} {\bibfnamefont {L.}~\bibnamefont {Calderin}},
  \ and\ \bibinfo {author} {\bibfnamefont {S.~B.}\ \bibnamefont {Trickey}},\
  }\bibfield  {title} {\enquote {\bibinfo {title} {Importance of
  finite-temperature exchange correlation for warm dense matter
  calculations},}\ }\href
  {https://journals.aps.org/pre/abstract/10.1103/PhysRevE.93.063207} {\bibfield
   {journal} {\bibinfo  {journal} {Phys. Rev. E}\ }\textbf {\bibinfo {volume}
  {93}},\ \bibinfo {pages} {063207} (\bibinfo {year} {2016})}\BibitemShut
  {NoStop}%
\bibitem [{\citenamefont {Mazevet}\ \emph {et~al.}(2005)\citenamefont
  {Mazevet}, \citenamefont {Desjarlais}, \citenamefont {Collins}, \citenamefont
  {Kress},\ and\ \citenamefont {Magee}}]{low_density1}%
  \BibitemOpen
  \bibfield  {author} {\bibinfo {author} {\bibfnamefont {S.}~\bibnamefont
  {Mazevet}}, \bibinfo {author} {\bibfnamefont {M.~P.}\ \bibnamefont
  {Desjarlais}}, \bibinfo {author} {\bibfnamefont {L.~A.}\ \bibnamefont
  {Collins}}, \bibinfo {author} {\bibfnamefont {J.~D.}\ \bibnamefont {Kress}},
  \ and\ \bibinfo {author} {\bibfnamefont {N.~H.}\ \bibnamefont {Magee}},\
  }\bibfield  {title} {\enquote {\bibinfo {title} {Simulations of the optical
  properties of warm dense aluminum},}\ }\href
  {https://journals.aps.org/pre/abstract/10.1103/PhysRevE.71.016409} {\bibfield
   {journal} {\bibinfo  {journal} {Phys. Rev. E}\ }\textbf {\bibinfo {volume}
  {71}},\ \bibinfo {pages} {016409} (\bibinfo {year} {2005})}\BibitemShut
  {NoStop}%
\bibitem [{\citenamefont {Desjarlais}\ \emph {et~al.}(2002)\citenamefont
  {Desjarlais}, \citenamefont {Kress},\ and\ \citenamefont
  {Collins}}]{low_density2}%
  \BibitemOpen
  \bibfield  {author} {\bibinfo {author} {\bibfnamefont {M.~P.}\ \bibnamefont
  {Desjarlais}}, \bibinfo {author} {\bibfnamefont {J.~D.}\ \bibnamefont
  {Kress}}, \ and\ \bibinfo {author} {\bibfnamefont {L.~A.}\ \bibnamefont
  {Collins}},\ }\bibfield  {title} {\enquote {\bibinfo {title} {Electrical
  conductivity for warm, dense aluminum plasmas and liquids},}\ }\href
  {https://journals.aps.org/pre/abstract/10.1103/PhysRevE.66.025401} {\bibfield
   {journal} {\bibinfo  {journal} {Phys. Rev. E}\ }\textbf {\bibinfo {volume}
  {66}},\ \bibinfo {pages} {025401(R)} (\bibinfo {year} {2002})}\BibitemShut
  {NoStop}%
\bibitem [{\citenamefont {Groth}\ \emph
  {et~al.}(2017{\natexlab{b}})\citenamefont {Groth}, \citenamefont {Dornheim},
  \citenamefont {Sjostrom}, \citenamefont {Malone}, \citenamefont {Foulkes},\
  and\ \citenamefont {Bonitz}}]{groth_prl}%
  \BibitemOpen
  \bibfield  {author} {\bibinfo {author} {\bibfnamefont {S.}~\bibnamefont
  {Groth}}, \bibinfo {author} {\bibfnamefont {T.}~\bibnamefont {Dornheim}},
  \bibinfo {author} {\bibfnamefont {T.}~\bibnamefont {Sjostrom}}, \bibinfo
  {author} {\bibfnamefont {F.~D.}\ \bibnamefont {Malone}}, \bibinfo {author}
  {\bibfnamefont {W.~M.~C.}\ \bibnamefont {Foulkes}}, \ and\ \bibinfo {author}
  {\bibfnamefont {M.}~\bibnamefont {Bonitz}},\ }\bibfield  {title} {\enquote
  {\bibinfo {title} {Ab initio exchange--correlation free energy of the uniform
  electron gas at warm dense matter conditions},}\ }\href
  {https://journals.aps.org/prl/abstract/10.1103/PhysRevLett.119.135001}
  {\bibfield  {journal} {\bibinfo  {journal} {Phys. Rev. Lett.}\ }\textbf
  {\bibinfo {volume} {119}},\ \bibinfo {pages} {135001} (\bibinfo {year}
  {2017}{\natexlab{b}})}\BibitemShut {NoStop}%
\bibitem [{\citenamefont {Ramakrishna}\ \emph {et~al.}()\citenamefont
  {Ramakrishna}, \citenamefont {Dornheim},\ and\ \citenamefont
  {Vorberger}}]{kushal}%
  \BibitemOpen
  \bibfield  {author} {\bibinfo {author} {\bibfnamefont {K.}~\bibnamefont
  {Ramakrishna}}, \bibinfo {author} {\bibfnamefont {T.}~\bibnamefont
  {Dornheim}}, \ and\ \bibinfo {author} {\bibfnamefont {J.}~\bibnamefont
  {Vorberger}},\ }\href {https://arxiv.org/abs/2002.11574} {}\ (\bibinfo
  {publisher} {Influence of finite temperature Exchange-Correlation effects in
  Hydrogen})\ \Eprint {http://arxiv.org/abs/2002.11574} {arXiv:2002.11574}
  \BibitemShut {NoStop}%
\bibitem [{ene()}]{energy_note}%
  \BibitemOpen
  \href@noop {} {}\bibinfo {note} {In fact, the kinetic and interaction energy
  of the unperturbed system scales (in first order) as $K\sim r_s^{-2}$ and
  $V\sim r_s^{-1}$, respectively. Yet, defining a rescaled perturbation
  amplitude [e.g., $\delta=A/(K+V)$] does not significantly
  simplify}\BibitemShut {NoStop}%
\bibitem [{\citenamefont {Bloembergen}(1982)}]{bloembergen_rmp_82}%
  \BibitemOpen
  \bibfield  {author} {\bibinfo {author} {\bibfnamefont {N.}~\bibnamefont
  {Bloembergen}},\ }\bibfield  {title} {\enquote {\bibinfo {title} {Nonlinear
  optics and spectroscopy},}\ }\href
  {https://journals.aps.org/rmp/abstract/10.1103/RevModPhys.54.685} {\bibfield
  {journal} {\bibinfo  {journal} {Rev. Mod. Phys}\ }\textbf {\bibinfo {volume}
  {54}},\ \bibinfo {pages} {685--695} (\bibinfo {year} {1982})}\BibitemShut
  {NoStop}%
\bibitem [{\citenamefont {Mukamel}\ and\ \citenamefont
  {Loring}(1986)}]{mukamel_86}%
  \BibitemOpen
  \bibfield  {author} {\bibinfo {author} {\bibfnamefont {S.}~\bibnamefont
  {Mukamel}}\ and\ \bibinfo {author} {\bibfnamefont {R.~F.}\ \bibnamefont
  {Loring}},\ }\bibfield  {title} {\enquote {\bibinfo {title} {Nonlinear
  response function for time-domain and frequency-domain four-wave mixing},}\
  }\href {https://www.osapublishing.org/josab/abstract.cfm?uri=josab-3-4-595}
  {\bibfield  {journal} {\bibinfo  {journal} {J. Opt. Soc. Am. B}\ }\textbf
  {\bibinfo {volume} {3}},\ \bibinfo {pages} {595--606} (\bibinfo {year}
  {1986})}\BibitemShut {NoStop}%
\bibitem [{\citenamefont {Axt}\ and\ \citenamefont
  {Mukamel}(1998)}]{axt_rmp_98}%
  \BibitemOpen
  \bibfield  {author} {\bibinfo {author} {\bibfnamefont {V.~M.}\ \bibnamefont
  {Axt}}\ and\ \bibinfo {author} {\bibfnamefont {S.}~\bibnamefont {Mukamel}},\
  }\bibfield  {title} {\enquote {\bibinfo {title} {Nonlinear optics of
  semiconductor and molecular nanostructures; a common perspective},}\ }\href
  {https://journals.aps.org/rmp/abstract/10.1103/RevModPhys.70.145} {\bibfield
  {journal} {\bibinfo  {journal} {Rev. Mod. Phys}\ }\textbf {\bibinfo {volume}
  {70}},\ \bibinfo {pages} {145--174} (\bibinfo {year} {1998})}\BibitemShut
  {NoStop}%
\bibitem [{\citenamefont {Sch\"utte}\ \emph {et~al.}(2012)\citenamefont
  {Sch\"utte}, \citenamefont {Bauch}, \citenamefont {Fr\"uhling}, \citenamefont
  {Wieland}, \citenamefont {Gensch}, \citenamefont {Pl\"onjes}, \citenamefont
  {Gaumnitz}, \citenamefont {Azima}, \citenamefont {Bonitz},\ and\
  \citenamefont {Drescher}}]{schuette_prl_12}%
  \BibitemOpen
  \bibfield  {author} {\bibinfo {author} {\bibfnamefont {B.}~\bibnamefont
  {Sch\"utte}}, \bibinfo {author} {\bibfnamefont {S.}~\bibnamefont {Bauch}},
  \bibinfo {author} {\bibfnamefont {U.}~\bibnamefont {Fr\"uhling}}, \bibinfo
  {author} {\bibfnamefont {M.}~\bibnamefont {Wieland}}, \bibinfo {author}
  {\bibfnamefont {M.}~\bibnamefont {Gensch}}, \bibinfo {author} {\bibfnamefont
  {E.}~\bibnamefont {Pl\"onjes}}, \bibinfo {author} {\bibfnamefont
  {T.}~\bibnamefont {Gaumnitz}}, \bibinfo {author} {\bibfnamefont
  {A.}~\bibnamefont {Azima}}, \bibinfo {author} {\bibfnamefont
  {M.}~\bibnamefont {Bonitz}}, \ and\ \bibinfo {author} {\bibfnamefont
  {M.}~\bibnamefont {Drescher}},\ }\bibfield  {title} {\enquote {\bibinfo
  {title} {Evidence for chirped auger-electron emission},}\ }\href
  {https://journals.aps.org/prl/abstract/10.1103/PhysRevLett.108.253003}
  {\bibfield  {journal} {\bibinfo  {journal} {Phys. Rev. Lett}\ }\textbf
  {\bibinfo {volume} {108}},\ \bibinfo {pages} {253003} (\bibinfo {year}
  {2012})}\BibitemShut {NoStop}%
\bibitem [{\citenamefont {Karasiev}\ \emph {et~al.}(2019)\citenamefont
  {Karasiev}, \citenamefont {Trickey},\ and\ \citenamefont {Dufty}}]{status}%
  \BibitemOpen
  \bibfield  {author} {\bibinfo {author} {\bibfnamefont {V.~V.}\ \bibnamefont
  {Karasiev}}, \bibinfo {author} {\bibfnamefont {S.~B.}\ \bibnamefont
  {Trickey}}, \ and\ \bibinfo {author} {\bibfnamefont {J.~W.}\ \bibnamefont
  {Dufty}},\ }\bibfield  {title} {\enquote {\bibinfo {title} {Status of
  free-energy representations for the homogeneous electron gas},}\ }\href
  {https://journals.aps.org/prb/abstract/10.1103/PhysRevB.99.195134} {\bibfield
   {journal} {\bibinfo  {journal} {Phys. Rev. B}\ }\textbf {\bibinfo {volume}
  {99}},\ \bibinfo {pages} {195134} (\bibinfo {year} {2019})}\BibitemShut
  {NoStop}%
\end{thebibliography}%

\end{document}